\documentclass[authoryear,preprint,review,12pt]{elsarticle}

\usepackage{graphicx}
\usepackage[table]{xcolor} 
\graphicspath{ {./pics/} }
\usepackage[official]{eurosym} 
\usepackage{hyperref} 
\hypersetup{
    colorlinks=true,
    linkcolor=blue,
    filecolor=magenta,      
    urlcolor=cyan,
    }
\usepackage{rotating}
\usepackage{tikz}
\usepackage{caption}
\usepackage[labelformat=simple]{subcaption}

\usepackage{amssymb}
\usepackage{amsthm}
\usepackage{amsmath}

\everymath{\displaystyle}

\usepackage{float}
\usepackage{tablefootnote} 

\usepackage{nomencl}

\newlength{\nomitemorigsep}
\makenomenclature
\usepackage{etoolbox}
\renewcommand\nomgroup[1]{%
  \item[\bfseries
  \ifstrequal{#1}{A}{Indices}{%
  \ifstrequal{#1}{B}{Sets}{%
  \ifstrequal{#1}{C}{Parameters}{%
  \ifstrequal{#1}{D}{Variables}{}}}}%
]}

\journal{Energy Policy}

\begin{document}

\begin{frontmatter}



\title{Power-to-X in Energy Hubs: Operations and Policies Supporting the Scale-Up of Renewable Fuel Production}


\author[inst1]{Ioannis Kountouris\corref{cor1}}

\cortext[cor1]{Corresponding author}
\ead{iokoun@dtu.dk}

\affiliation[inst1]{organization={Technical University of Denmark},
            addressline={Department of Technology, Management and Economics, Produktionstorvet, Building 424}, 
            city={Kongens Lyngby},
            postcode={2800}, 
            country={Denmark}}

\author[inst1]{Lissy Langer}
\author[inst1]{Rasmus Bramstoft}
\author[inst1]{Marie M\"unster}
\author[inst1]{Dogan Keles}

\begin{abstract}
Power-to-X (P2X) needs to scale up rapidly to provide the fuels required in the hard-to-decarbonize industrial and heavy transport sector. Only recently, the European Commission proposed requirements for \textit{renewable} fuels. P2X energy hubs enable efficient synergies between energy infrastructures, production facilities, and storage options. In this study, we explore the optimal operation of an energy hub by leveraging the flexibility of P2X including hydrogen, methanol, and ammonia synthesizers, and analyze potential revenue streams such as the day-ahead and ancillary service markets. We propose EnerHub2X, a mixed-integer linear program that maximizes the hub's profit based on current market prices, considering technical constraints of P2X such as unit commitment and non-linear efficiencies. We model a representative Danish energy hub and find that without price incentives, it mainly produces liquid hydrogen and sells renewable electricity. Only by adding a price premium of about 50\% (0.16 \euro{}/kg) to the conventional fuel prices, sufficient amounts of renewable ammonia and methanol are produced. To utilize production efficiently, on-site renewable capacity and P2X must be carefully aligned. We show that renewable power purchase agreements can provide flexibility while complying with the rules set by the European Commission.
\end{abstract}



\begin{highlights}
 
%
%

\item We propose a new multi-commodity flow model to analyze Power-to-X in energy hubs.
\item We explore optimal operation for different expansion pathways and markets.
\item We evaluate the profitability of operating energy hubs under current market conditions.
\item We analyze different policy instruments promoting renewable fuel production. 
\item We illustrate the dispatch merit order for varying levels of renewable fuel premium.

\end{highlights}


\end{frontmatter}


\section{Introduction} \label{sec:intro}
The climate targets of the European Commission (EC) require a redesign of future energy systems due to the stochastic nature of renewable energy sources~(RES) and limited sustainable carbon sources. The integration of Power-to-X~(P2X) in a renewable-based energy system will be crucial, especially for the decarbonization of hard-to-abate sectors such as industry and heavy transport. In addition, P2X enables seasonal storage, provides system flexibility, and allows integration across energy vectors (i.e., power, gas, and heat). 
To scale up P2X production in the near future, energy hubs play an important role by demonstrating viable business models and establishing operations on market-based terms \citep{energinet_winds_2019}. 

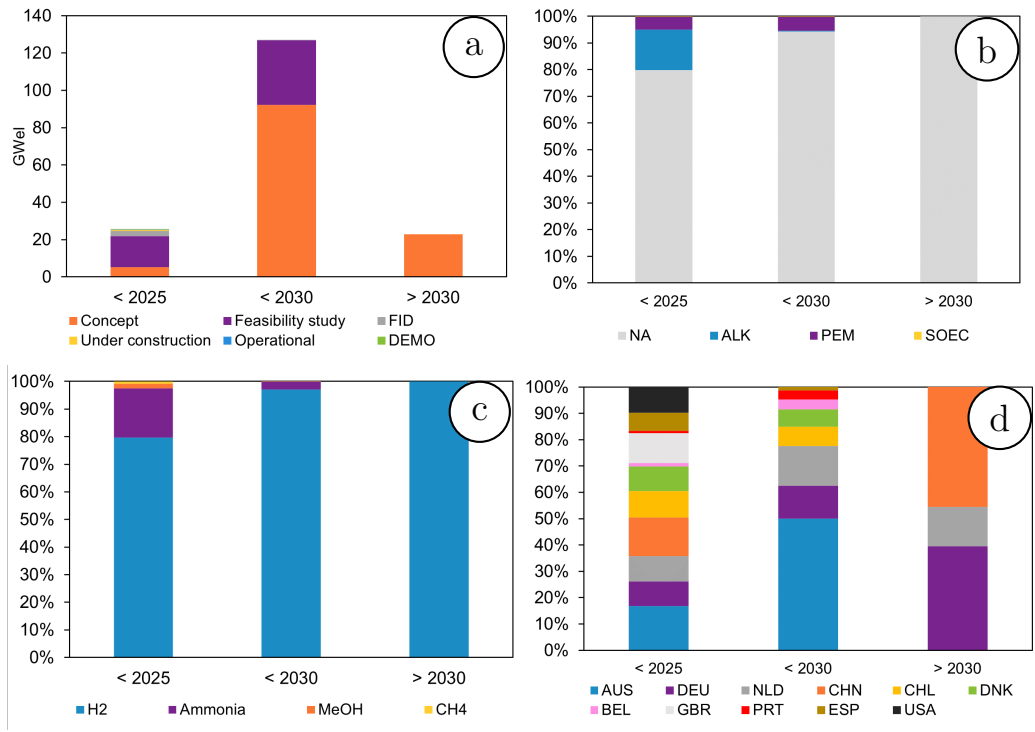
\begin{figure}[ht]
    \centering
    \begin{subfigure}[b]{0.5\linewidth}
        \centering
        \begin{tikzpicture} 
            \node[anchor=south west,inner sep=0] (imagea) at (0,0) {\includegraphics[width=\textwidth]{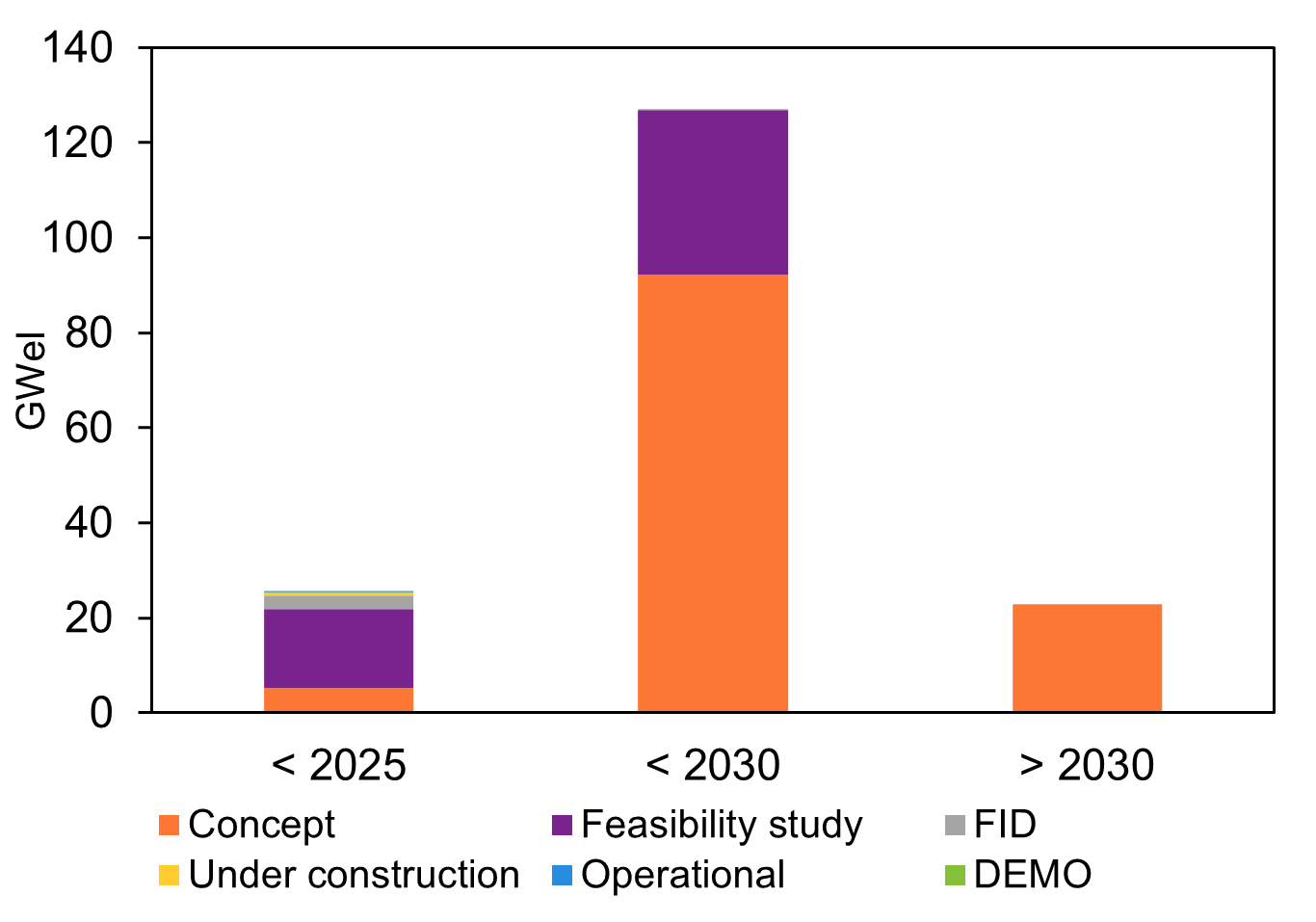}};
            \node (a) at (6.21,4.15) [thick, draw=black, circle, fill=white, minimum size=.8cm] {\large{a}}; 
        \end{tikzpicture}
        \phantomsubcaption
        \label{fig:h2-state}
    \end{subfigure}%
    \hfill
    \begin{subfigure}[b]{0.5\linewidth}
        \centering
        \begin{tikzpicture} 
            \node[anchor=south west,inner sep=0] (imageb) at (0,0) {\includegraphics[width=\textwidth]{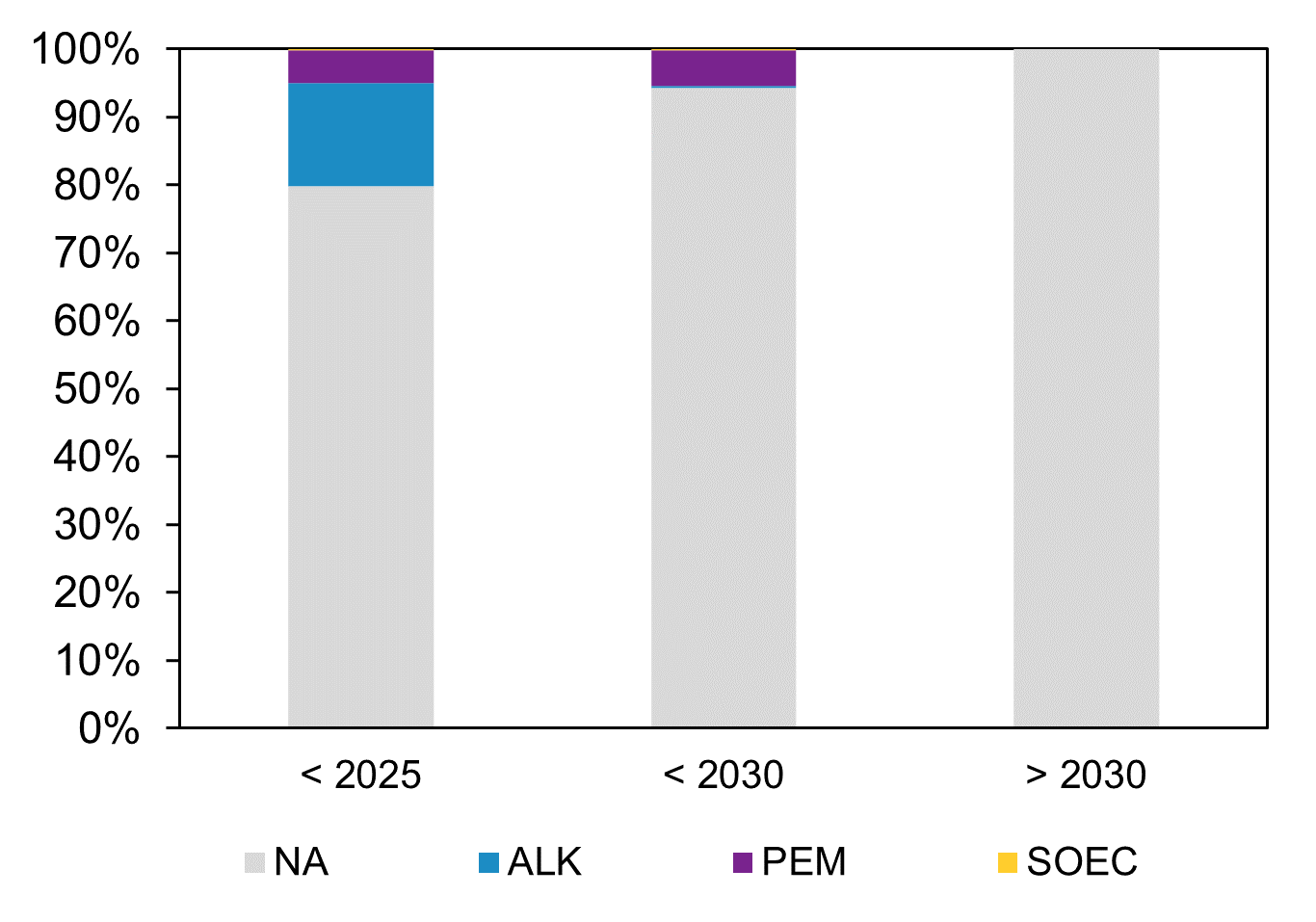}};
            \node (b) at (6.18,4.12) [thick, draw=black, circle, fill=white, minimum size=.8cm] {\large{b}}; 
        \end{tikzpicture}
        \phantomsubcaption
        \label{fig:h2-tech}
    \end{subfigure}
    
    \begin{subfigure}[b]{0.5\textwidth}
    \vspace{-.7cm} 
        \centering
        \begin{tikzpicture} 
            \node[anchor=south west,inner sep=0] (imaged) at (0,0) {\includegraphics[width=\textwidth]{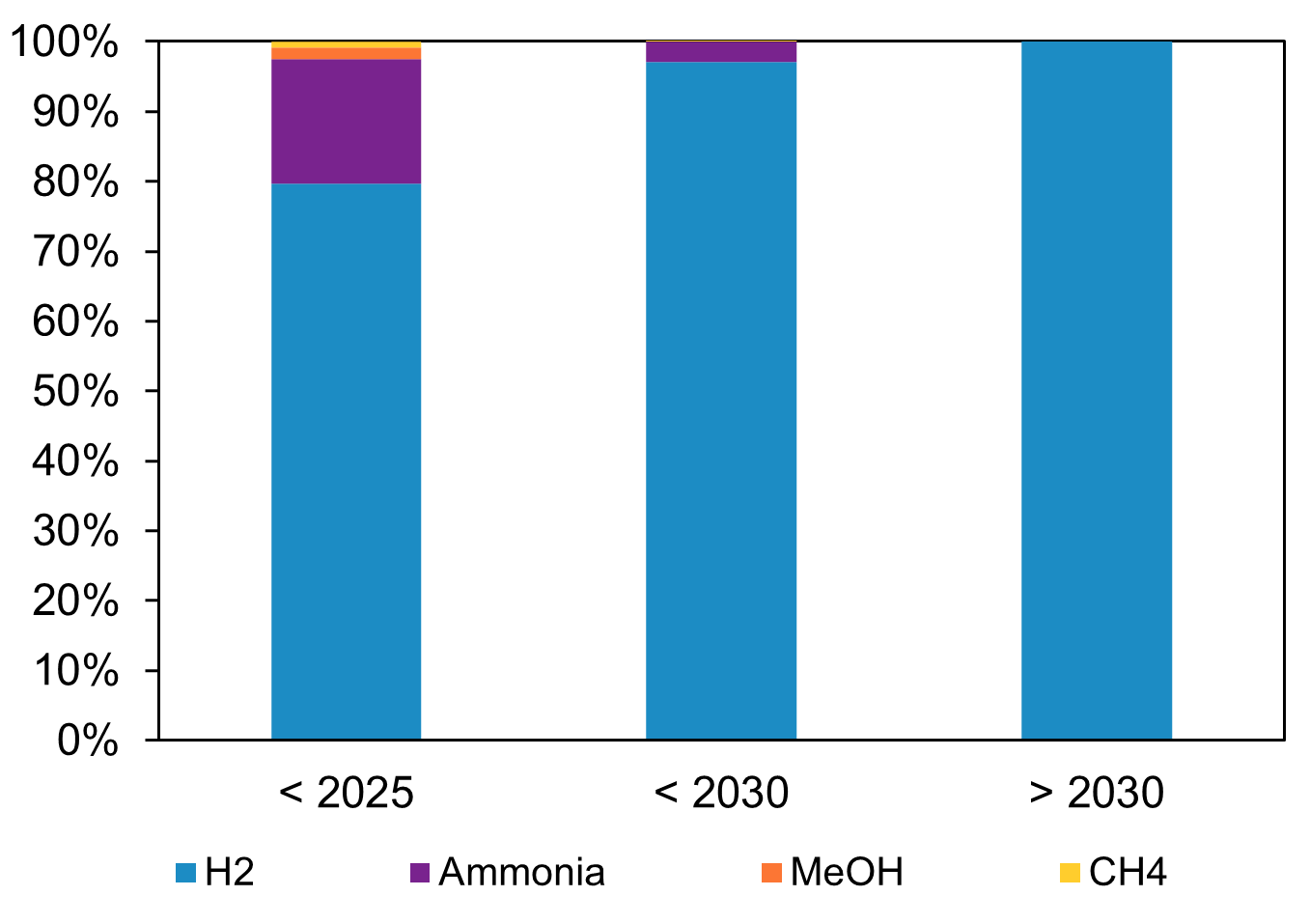}};
        \node (c) at (6.29,4.23) [thick, draw=black, circle, fill=white, minimum size=.8cm] {\large{c}}; 
        \end{tikzpicture}
        \phantomsubcaption
        \label{fig:h2-output}
    \end{subfigure}%
    \hfill
    \begin{subfigure}[b]{.5\textwidth}
    \vspace{-.7cm} 
        \centering
        \begin{tikzpicture} 
            \node[anchor=south west,inner sep=0] (imagec) at (0,0) {\includegraphics[width=\textwidth]{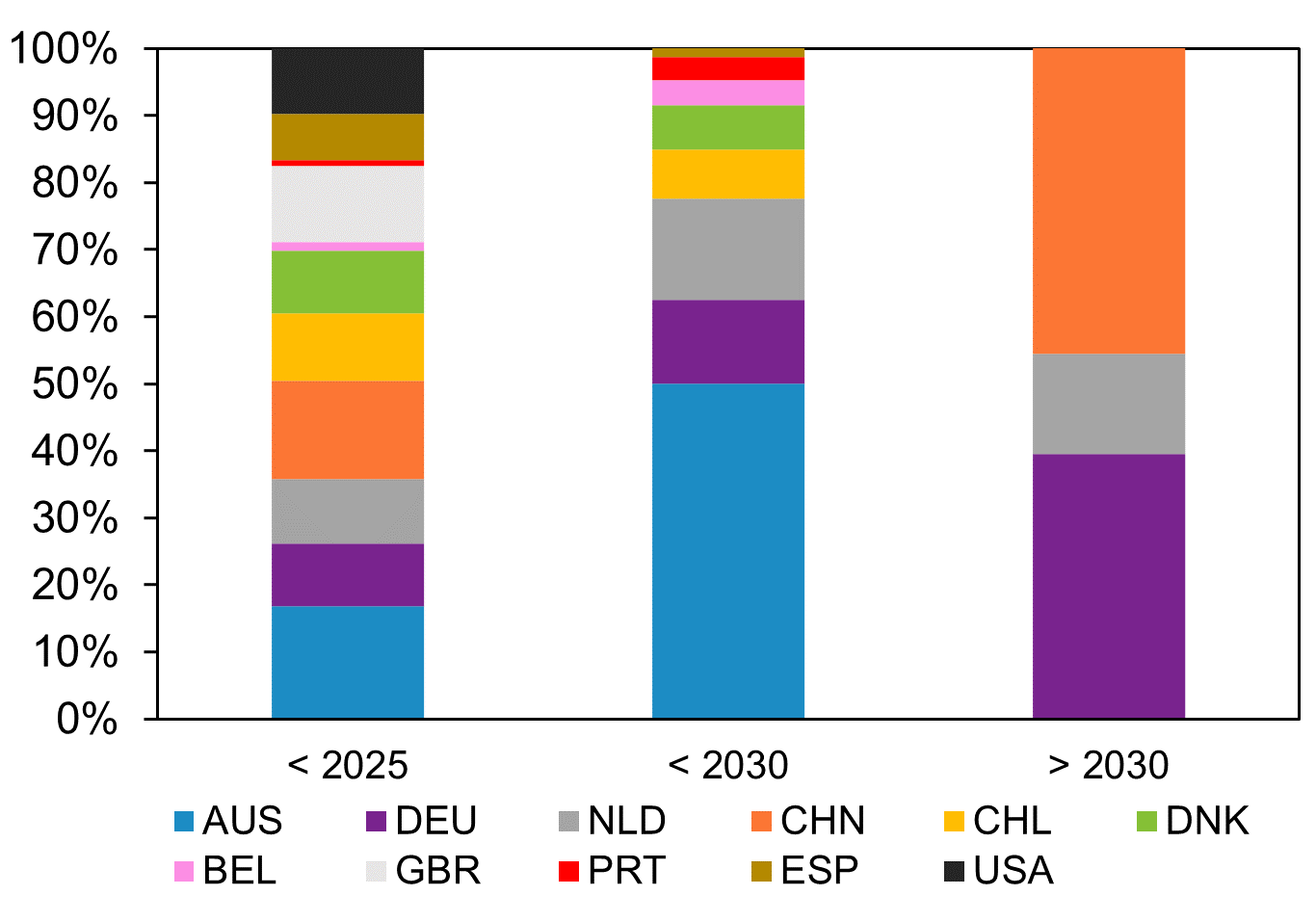}}; 
        \node (d) at (6.35,4.15) [thick, draw=black, circle, fill=white, minimum size=.8cm] {\large{d}}; 
        \end{tikzpicture}
        \phantomsubcaption
        \label{fig:h2-country}
    \end{subfigure}
    \vspace{-1cm} 
    \caption{IEA Database of global hydrogen projects: a) Project status, b) Electrolyzer technology, c) Final product, d) Project country (data: \citet{iea_hydrogen_2021})}
\end{figure}

The REPowerEU initiative of the \citet{european_commission_repowereu_2022} sets the goal of 10~Mt of annual domestic production and 10~Mt of imports of renewable hydrogen by 2030. This requires an ambitious European electrolyzer capacity of at least 64~GW\textsubscript{e}\footnote{This assumes 100\% utilization and the (current) technology conversion factors used by the \citet{iea_hydrogen_2021}: 0.0046 (ALK), 0.0052 (PEM) MW/nm\textsuperscript{3} H\textsubscript{2}/hour and a hydrogen density of 0.089 kg/m\textsuperscript{3}.}. The market environment, however, is highly uncertain. Based on the hydrogen project database of the \citet{iea_hydrogen_2021}, Figure~\ref{fig:h2-state} shows that most global hydrogen projects are still in their concept phase, and almost all projects have not yet decided which technology to use (Figure~\ref{fig:h2-tech}) and which fuel to produce (Figure~\ref{fig:h2-output}). Currently, a lot of different countries\footnote{Illustrated are the eleven countries with project capacities of more than 1.5 GW\textsubscript{e}} are planning pilot projects in the coming years, some with giga-scale ambitions (Figure~\ref{fig:h2-country}). To fulfill the EC's goals as well as target the 850 GW\textsubscript{e} of global electrolyzer capacity needed according to the \textit{Net Zero Emissions by 2050 Scenario} of the \citet{iea_net_2021}, large-scale projects need to ramp up soon. Yet, until 2022, only five projects with a capacity greater than 10 MW\textsubscript{e} have become operational \citep{iea_hydrogen_2021}. Table \ref{tab:h2-projects} describes these and some additional projects projected to go live until 2025 with a capacity of more than 10 MW\textsubscript{e}. Mostly, these are initiated by large utility companies in cooperation with an international mix of technology providers, varying electrolyzer technologies, and final products. Yet, all of these have in common, that they are connected to regional energy hubs such as mobility stations (using fuel cells), ports, upgraded refineries, or ammonia production sites as these enable efficient synergies between energy infrastructures, production facilities, and storage options. In most of these projects, the capacity is already defined, companies seem, however, reluctant to decide on a technology, a provider, and a fuel to produce. Despite recent political ambitions, the lack of clear policy support seems to stall industrial commitments. 

In addition, market participants still wait for a clear definition of what is considered \textit{renewable}. Finally, in May 2022, the \citet{european_commission_supplementary_2022} proposed a delegated act for renewable fuel production requirements considering additionality and the temporal and geographic correlation of the RES used (see further in Section \ref{sec:case_2}). These rules will probably also apply to renewable hydrogen used in industry and for imports. The legislation is expected to be updated based on the submitted public feedback and to be introduced in the upcoming RED II \citep{european_commission_directive_2018} update in autumn 2022.

In conclusion, we note that additional support is needed for market participants investing in such P2X energy hubs, but also for policy makers trying to incentivize renewable fuel production. This should consider current market conditions such as prices and regulations, as well as relevant technical details of P2X plants and potential synergies within energy hubs.\\

Hence, the goal of this study is to investigate whether regional energy hubs with renewable fuel production can establish economically viable operations under current market conditions. In addition, we analyze which support schemes can be utilized to promote renewable fuels taking into account the recent EC's production requirements. In order to do that:

\begin{itemize}
  \item We introduce a flexible modeling framework for energy hubs focusing on renewable fuels and cluster synergies.
  \item We analyze optimal flows of an expansion pathway and investigate potential revenue streams, including ancillary service markets.
  \item We investigate different policy instruments to promote renewable fuel production in a Danish case study.
\end{itemize}

The remainder of this paper is structured as follows: Section \ref{sec:related-work} presents related work on P2X energy hubs. Section \ref{sec:model-description} introduces the main components of the proposed EnerHub2X model, whereas the whole model formulation and the used data can be found in \ref{sec:appendix:MathematicalModel} and \ref{sec:appendix:skive-data}, respectively. Section \ref{sec:skive-case-study} introduces the Danish case study and Section \ref{sec:numerical-analysis} the numerical analysis. Section \ref{sec:results} discusses the results, limitations, and further research and Section \ref{sec:conclusion} concludes with the policy recommendations.

\section{Literature Review} \label{sec:related-work}
An energy hub (EH) is defined as a multi-energy system, where various energy carriers are optimally produced, converted, stored, and consumed \citep{mohammadi_energy_2017}, fulfilling certain sociopolitical and socioeconomic mandates \citep{ding_review_2022}. The concept was introduced by \citet{geidl_optimal_2007} using a non-linear and a linearized model (with constant plant efficiencies) to optimally dispatch electricity, natural gas, and district heating using different EHs with combined heat and power (CHP) units. \citet{almassalkhi_optimization_2011} decompose the framework into matrices, but despite this structure, changing network topologies is still quite costly. In addition, all these efforts only consider a single input per conversion unit. For example, \citet{hajimiragha_optimal_2007} and \citet{krause_multiple-energy_2011} in their \textit{hydrogen economy} application use only electricity as input to their electrolyzer, which, for example, limits applicability in countries with rising water scarcity. To facilitate synergy networks using excess products (e.g., wastewater), an integrated management framework is required that considers multi-source multi-product converters \citep{hemmes_towards_2007,mohammadi_energy_2017}. These multi-input/multi-output systems contain several energy resources, converters, transmission lines, storage systems, and loads \citep{aljabery_multi_2021}. They aim at improving reliability, flexibility, and efficiency by coupling, coordinating, and co-optimizing multiple forms of energy, such as electricity, heating and cooling, as well as e-fuels \citep{chen_analyzing_2018} towards a 100\% renewable energy system \citep{ghasemi_integrated_2018,ding_review_2022, mohammadi_energy_2017}. 

Research on \textit{multi-carrier energy systems} or \textit{multi-energy systems}, \textit{energy hubs}, or \textit{integrated energy systems} has been conducted extensively for more than a decade now. A  multitude of review studies have been published focusing on topology \citep{son_multi_2021, mancarella_mes_2014}, optimization techniques for planning, operation, and trading \citep{ding_review_2022,aljabery_multi_2021}, focus of the analysis \citep{mohammadi_energy_2017,sadeghi_energy_2019}, or the current  available infrastructure \citep{guelpa_towards_2019}. 

The dominant conversion technologies investigated are, however, still CHP plants and gas boilers, the dominant energy carriers are electricity and natural gas, and the dominant demands are electricity and heating \citep{mohammadi_energy_2017, son_multi_2021, sadeghi_energy_2019}. Studies mainly consider electrolyzers to consume electricity in combination with CHP or fuel cells. Yet, as a consequence of the recent increase in urgency to reduce fossil fuel dependencies, including hard-to-decarbonize sectors, the interest in how to model and optimize these sector-coupled systems has rejuvenated. Hence, we focus our investigation on EHs using renewable hydrogen to produce e-fuels that can be used in the industry or heavy transport sectors.

Multiple studies investigate separately the production of hydrogen in combination with fuel cell electric vehicles \citep{nasir_optimal_2022,maroufmashat_mixed_2016}, comparing the economies of different production technologies \citep{maniyali_energy_2013, sharif_design_2014,parra_integrated_2017} or different electricity markets \citep{luck_economic_2017,scolaro_optimizing_2022}, the production of methanol \citep{chen_power--methanol_2021, zheng_data-driven_2022, zhang_integrated_2019} in a Icelandic/German case study \citep{kauw_green_2015}, or the production of ammonia \citep{salmon_green_2021, zhang_techno-economic_2020} in case studies in the UAE \citep{osman_scaling_2020}, South America \citep{armijo_flexible_2020}, or the USA \citep{morgan_sustainable_2017}.  Overarching, \citet{decourt_weaknesses_2019} analyzes the P2X innovation system, and \citet{chehade_review_2019} and \citet{gahleitner_hydrogen_2013} review global pilot projects. \citet{bellotti_comparative_2022} compare the economic feasibility of different renewable P2X plants and find that in the close-to-mid-term time horizon, electrical energy should be stored in renewable fuel to be transported using existing infrastructure and to circumvent issues related to hydrogen transportation.

Utilizing an integrated energy system modelling approach, \citet{bramstoft_modelling_2020} assess renewable gas and liquid fuel production pathways for the hard-to-abate sectors considering synergies and interactions between energy sectors and vectors in a Danish case study. \citet{lester_analysis_2020} build upon this methodology and perform a long-term analysis of an extensive catalog of various P2X and e-fuel production pathways. They find that e-methanol and e-ammonia are promising options to supply the long-haul transport sector and the maritime sector. 
\citet{incer-valverde_hydrogen-driven_2022} compare different P2X fuels in a multi-criteria evaluation of a German case study. They find ammonia to be the best option, closely followed by methanol. Finally, in their techno-economic cradle-to-grave environmental assessment, \citet{ordonez_carbon_2021} find similar net production costs for renewable methanol and ammonia and stress the high environmental impact of H\textsubscript{2} storage. 

Overall, a techno-economic analysis of the operation of a synergetic energy hub producing renewable hydrogen and e-fuels, as well as of the necessary policy frameworks to incentivize renewable fuel production, have to the best of our knowledge, not been conducted yet. Therefore, in this study, we develop a state-of-the-art energy hub operation model called EnerHub2X. EnerHub2X allows optimizing small network flow problems but also generalizes to large-scale P2X value chains operating under a common energy hub framework. The flexible formulation allows modeling of additional conversion technologies and easy adoption to different cluster topologies.

\section{Methodology: EnerHub2X} \label{sec:model-description}
EnerHub2X is designed to capture the complexity and dynamics of P2X and provides insights into optimal dispatch and operational bottlenecks. The model flexibly simulates different processes and conversion technologies using accurate energy carrier transformation including technical and operational constraints. EnerHub2X is a network and multi-commodity flow model solved in a four-dimensional space: accounting for spatial and temporal resolution, technology details, and energy carriers. Its objective is to maximize the energy hub's profit and optimally dispatch all its conversion technologies assuming perfect information. The planning horizon can be either daily, weekly, or yearly using an hourly time resolution. The model assumes a connection to the electricity grid, the day-ahead market, and the ancillary service market.

EnerHub2X is formulated as a mixed-integer linear program in GAMS and solved using CPLEX. In the following section, we introduce the model formulation of EnerHub2X denoting all parameter names in capital (except ratios and fuel prices) and variable names in small letters. The nomenclature and complete mathematical formulation including the ancillary service market can be found in \ref{sec:appendix:MathematicalModel}.

\subsection{Objective Function}
EnerHub2X maximizes the energy hub's profit (Eq. \ref{eq: ObjectivefunctionderPro}) composed of four terms: the market price $\pi^{sale}_{a,e,t}$ multiplied by the quantity $f^{sale}_{a,e,t}$ of each fuel~$e$ sold in area $a$ in time period $t$ as the energy hub's revenue from fuel sales; the market price $\pi^{buy}_{a,e,t}$ multiplied by the feed-stock purchased~$f^{buy}_{a,e,t}$ as the fuel costs; the variable operational costs $C^{var}_{g}$ of technology $g$ multiplied with the produced quantities $x^{out}_{g,e,t}$; and the ramp-up costs $c^{start}_{g,t}$ for technologies with unit commitment $g\in\mathbb{G}^{UC}$. 

\begin{equation}
\begin{aligned}
 \underset{
}{\operatorname{Maximize}}  
& \sum_{t}(\sum_{(a,e) \in \mathbb{E}^{sale}_{a}} \pi^{sale}_{a,e,t} f^{sale}_{a,e,t} - \sum_{(a,e) \in \mathbb{E}^{buy}_{a}}  \pi^{buy}_{a,e,t} f^{buy}_{a,e,t}   
\\&  
- \sum_{(g,e) \in \mathbb{E}^{def}_{g}} C^{var}_{g} x^{out}_{g,e,t}  - \sum_{g \in \mathbb{G}^{UC}} c^{start}_{g,t})
\label{eq: ObjectivefunctionderPro}
\end{aligned}
\end{equation}

\subsection{Fuel Mix and Conversion}
EnerHub2X's core functionality is the ability to model conversion technologies based on input $F^{in}_{g,e}$ and output $F^{out}_{g,e}$ flows. These parameters capture the flow mass of each technology $g$ and each fuel $e$. Based on those parameters the values $\sigma^{in}_{g,e},\sigma^{out}_{g,e}$, and $\Theta_{g}$ are determined. $\sigma^{in}_{g,e}$ describes the input share of fuel $e$ used in technology $g$ (Eq. \ref{eq:Insharedefinied}), $\sigma^{out}_{g,e}$ describes the output share of fuel $e$ of technology $g$ (Eq. \ref{eq:Outsharedefinied}), and $\Theta_{g}$ describes the ratio between the sum of input flows and the sum of output flows of technology $g$ (Eq. \ref{eq:PercentageDifference}).

\begin{equation}
    \sigma^{in}_{g,e}  = \frac{F^{in}_{g,e}}{ \sum\limits_{e \in \mathbb{E}^{in}_{g}}F^{in}_{g,e}} {E}^{in}_{g} \quad \forall g \in \mathbb{G} ,e \in \mathbb{E}^{in}_{g}
    \label{eq:Insharedefinied}
\end{equation}

\begin{equation}
    \sigma^{out}_{g,e}  = \frac{F^{out}_{g,e}}{ \sum\limits_{e \in \mathbb{E}^{out}_{g}}F^{out}_{g,e}}  \quad \forall g \in \mathbb{G} ,e \in \mathbb{E}^{out}_{g}
    \label{eq:Outsharedefinied}
\end{equation}

\begin{equation}
    \Theta_{g}  = \frac{\sum\limits_{e \in \mathbb{E}^{in}_{g}}F^{in}_{g,e}}{ \sum\limits_{e \in \mathbb{E}^{out}_{g}}F^{out}_{g,e}}  \quad \forall g \in \mathbb{G} 
    \label{eq:PercentageDifference}
\end{equation}

The decision variables $x^{in}_{g,e,t},x^{out}_{g,e,t}$ and $x^{total}_{g,t}$ describe the input and output flows, as well as the activity level of each technology $g$. Eq. \ref{eq:FuelUse} connects the consumption $x^{in}_{g,e,t}$ of fuel $e$ with the overall activity level $x^{total}_{g,t}$ of technology~$g$. Similarly, Eq. \ref{eq:Generation} links the generation $x^{out}_{g,e,t}$ of fuel $e$ with the total activity level. These linkages introduce linear conversion efficiencies.  

\begin{equation}
    x^{in}_{g,e,t}  =  \sigma^{in}_{g,e} x^{total}_{g,t}  \quad \forall g \in \mathbb{G} ,e \in \mathbb{E}, t \in \mathbb{T}
    \label{eq:FuelUse}
\end{equation}

\begin{equation}
    x^{out}_{g,e,t} = \sigma^{out}_{g,e} \Theta_{g} x^{total}_{g,t}    \quad \forall g \in  \mathbb{G} \backslash  ( \mathbb{G}^{S}  \cup \mathbb{G}^{LD} ) ,e \in \mathbb{E},t \in \mathbb{T}
    \label{eq:Generation}
\end{equation}

In addition, technologies with non-linear load-dependent production efficiencies $g \in \mathbb{G}^{LD}$ can be incorporated using SOS2 constraints (Special Ordered Sets of type 2) similar to \cite{lin_economic_2021} (Eqs.~\ref{eq:ImportFuelCurve} and \ref{eq:ProductionFuelCurve}). The breakpoints between each linear segment are denoted by $k$, while $\Theta_{g,k}$ and $U_{g,k}$ indicate the discrete efficiency and total level of activity of technology~$g \in \mathbb{G}^{LD} $ at breakpoint $k$. In addition, $P^{max}_{g}$ denotes the capacity of technology $g$ and $w^{SOS2}_{g,t,k}$ the weight assigned to the breakpoints. Using SOS2 constraints, all weights have to sum to one and only two subsequent weights can be larger or equal to zero.

\begin{equation}
     x^{total}_{g,t} =   (\sum_{k \in K}  w^{SOS2}_{g,t,k} U_{g,k}) P^{max}_{g} \quad \forall g \in \mathbb{G}^{LD} e \in \mathbb{E}, t \in \mathbb{T}
    \label{eq:ImportFuelCurve}
\end{equation}

\begin{equation}
     x^{out}_{g,e,t} =  \sigma^{out}_{g,e} (\sum_{k \in K} w^{SOS2}_{g,t,k} U_{g,k} \Theta_{g,k}) P^{max}_{g} \quad \forall g \in \mathbb{G}^{LD}  ,e \in \mathbb{E},t \in \mathbb{T}
    \label{eq:ProductionFuelCurve}
\end{equation}

For technologies $g \in \mathbb{G}^{LD}$ with unit commitment (i.e.,  $g \in \mathbb{G}^{UC}$), Eq.~\ref{eq:Weights} activates the unit commitment variable $o_{g,t}$ accordingly.

\begin{equation}
     \sum_{k \in K}  w^{SOS2}_{g,t,k} =  o_{g,t} \quad \forall g \in (\mathbb{G}^{LD} \cap  \mathbb{G}^{UC}) ,t \in \mathbb{T}
    \label{eq:Weights}
\end{equation}

\subsection{Technical Constraints}
The following constraints restrict technology operations (i.e., $x^{in}_{g,e,t},x^{out}_{g,e,t}$ and $x^{total}_{g,t})$ based on their respective characteristics. All technologies are linked to exogenous capacity factors $P^{profile}_{g,t} \in [0,1]$ (e.g., to force maintenance this parameter could be set to zero). Eq. \ref{eq:limitFuelusetot} restricts the total level of activity based on the capacity $P^{max}_{g}$ and $P^{profile}_{g,t}$. We assume all technologies are shut down initially.

\begin{equation}
       x^{total}_{g,t} \leq P^{max}_{g} P^{profile}_{g,t}  \quad \forall g,t \in \mathbb{T} 
\label{eq:limitFuelusetot}
\end{equation}

The flexibility of technology $g \in \mathbb{G}^{R} $ is restricted by its production rates $R^{up}_{g}$ and $R^{down}_{g}$ (Eqs.~\ref{eq:Ramp up no UC} and \ref{eq:Ramp Down no UC}).

\begin{equation}
       \sum_{e \in E} \left (x^{out}_{g,e,t} - x^{out}_{g,e,t-1} \right) \leq R^{up}_{g}  \quad \forall g \in  \mathbb{G}^{R} ,t \in \mathbb{T}
\label{eq:Ramp up no UC}
\end{equation}

\begin{equation}
       \sum_{e \in E} \left (x^{out}_{g,e,t-1} - x^{out}_{g,e,t} \right) \leq R^{down}_{g}  \quad \forall g \in  \mathbb{G}^{R} ,t \in T
\label{eq:Ramp Down no UC}
\end{equation}

Furthermore, the flexibility of some technologies $g \in (\mathbb{G}^{R} \cap \mathbb{G}^{UC})$ is further restricted by unit commitment constraints (Eq.~\ref{eq:maxminproduction}). The associated cost~$c^{start}_{g,t}$ is derived in Eq.~\ref{eq:Startupcost} considering the ramp-up decision $u_{g,t}$ (Eq.~\ref{eq:StartupcostCondition}).

\begin{equation}
      P^{min}_{g}o_{g,t}  \leq  x^{total}_{g,t}      \leq  P^{max}_{g}o^{}_{g,t}  \quad \forall g \in \mathbb{G}^{UC}, t \in \mathbb{T}
\label{eq:maxminproduction}
\end{equation}


\begin{equation}
        o_{g,t} - o_{g,t-1} \leq u_{g,t} \quad \forall g \in \mathbb{G}^{UC}, t \in \mathbb{T}
\label{eq:StartupcostCondition}
\end{equation}

\begin{equation}
       c^{start}_{g,t} = C^{start}_{g}u_{g,t}    \quad \forall g \in \mathbb{G}^{UC}, t \in \mathbb{T}
\label{eq:Startupcost}
\end{equation}

For unit commitment technologies, Equations~\ref{eq:Ramp up no UC} and \ref{eq:Ramp Down no UC} are expanded. In addition to the production rates, Eq. \ref{eq:Ramp Up UC} restricts the ramp-up rate to the minimum production level $P^{min}_{g}$ of a technology $g \in (\mathbb{G}^{R} \cap \mathbb{G}^{UC})$. This prevents shutting down a technology and forces it into a standby mode $P^{min}_{g}$ (Eq.~\ref{eq:maxminproduction}) to remain flexible. Similarly, Eq. \ref{eq:Ramp Down UC} restricts the ramp-down rate of a technology. 

\begin{equation}
\begin{aligned}
       \sum_{e \in E} \left (x^{out}_{g,e,t} - x^{out}_{g,e,t-1} \right) \leq R^{up}_{g} o_{g,t-1} + P^{min}_{g} (1-o_{g,t-1})  \quad \forall g \in (\mathbb{G}^{UC} \cap \mathbb{G}^{R}),t \in \mathbb{T}
\label{eq:Ramp Up UC}
\end{aligned}
\end{equation}

\begin{equation}
\begin{aligned}
       \sum_{e \in E} \left (x^{out}_{g,e,t-1} - x^{out}_{g,e,t} \right) \leq R^{down}_{g} o_{g,t} + P^{min}_{g} (1-o_{g,t})  \quad \forall g \in (\mathbb{G}^{UC} \cap \mathbb{G}^{R}),t \in \mathbb{T}
\label{eq:Ramp Down UC}
\end{aligned}
\end{equation}

\subsection{Storage Constraints}
EnerHub2X considers different types of storage technologies, e.g., battery electric or compressed hydrogen storage systems. The decision variable $soc_{g,t}$ represents the state of charge (SOC) of technology $g \in \mathbb{G}^{s}$ managed by the storage balance equation (Eq.~\ref{eq:ProductionStorage}). Eq.~\ref{eq:max volume} restricts the $soc_{g,t}$ to the maximum level $SOC^{max}_{g}$ and Eq.~\ref{eq:volinitLast} defines the initial values.

\begin{equation}
    soc_{g,t} = soc_{g,t-1} + \Theta_g x^{total}_{g,t} - \sum_{e \in E^{out}_{g}}  \sigma^{out}_{g,e} x^{out}_{g,e,t}   \quad \forall g \in \mathbb{G}^{s} ,e \in \mathbb{E}, t \in \mathbb{T}
    \label{eq:ProductionStorage}
\end{equation}

\begin{equation}
    soc_{g,t} \leq SOC^{max}_{g}    \quad \forall g \in \mathbb{G}^{s}, t \in \mathbb{T}
    \label{eq:max volume}
\end{equation}

\begin{equation}
    soc_{g,t=0} = soc_{g,t=T} = SOC^{init}_{g}   \quad \forall g \in \mathbb{G}^{s} 
    \label{eq:volinitLast}
\end{equation}

The binary decision variable $\nu_{g,t}$ prevents simultaneous charging and discharging. The maximum and minimum rates of dis-/charging are accounted for in Eqs. \ref{eq:max charging} and \ref{eq:max discharging}. 

\begin{equation}
  P^{min}_{g} \nu_{g,t}  \leq x^{total}_{g,t} \leq P^{max}_{g} \nu_{g,t}    \quad \forall g \in \mathbb{G}^{s}, t \in \mathbb{T}
    \label{eq:max charging}
\end{equation}


\begin{equation}
 P^{min}_{g} (1-\nu_{g,t})   \leq \sum_{e \in E^{out}_{g}}  \sigma^{out}_{g,e} x^{out}_{g,e,t}  \leq P^{max}_{g} (1-\nu_{g,t})    \quad \forall g \in \mathbb{G}^{s}, t \in \mathbb{T}
    \label{eq:max discharging}
\end{equation}



\subsection{Fuel Flow Balance and Restrictions}
The model connects different hierarchical areas using the fuel flows $\mathbb{F}_{(a,a\prime)}$ in the balance equation (Eq. \ref{eq:EnergyBalance}). All areas $a$ are connected with one another creating a network for each fuel $e$. Up to five decisions are made in each area. In areas with technologies $g \in G_a$, the model decides how much to consume and produce (i.e., $x^{in}_{g,e,t}$ and $x^{out}_{g,e,t}$) of fuel $e$. In addition, each area $a$ can purchase $q^{buy}_{a,e,t}$ or sell $q^{sale}_{a,e,t}$ fuel $e$. The im-/exporting decisions are linked with in-/out-flows $f^{flow}_{a,a\prime,e,t}$ of fuel $e$ to adjacent area $a\prime$.

\begin{equation}
\begin{aligned}
    & q^{buy}_{a,e,t} + \sum_{a' \in \mathbb{F}_{(a\prime,a)}} f^{flow}_{a\prime,a,e,t} + \sum_{g \in (\mathbb{G}_{a} \cap \mathbb{E}^{out}_{g})} x^{out}_{g,e,t} =     
   \\
   & q^{sale}_{a,e,t} + \sum_{a \in \mathbb{F}_{(a,a\prime)}} f^{flow}_{a,a\prime,e,t} + \sum_{g \in (\mathbb{G}_{a} \cap \mathbb{E}^{in}_{g})} x^{in}_{g,e,t}            \quad \forall  (a,a') \in \mathbb{A}, e \in \mathbb{E},t \in \mathbb{T}
    \label{eq:EnergyBalance}
\end{aligned}
\end{equation}

Overall, the EnerHub2X framework can be used to model energy hubs of varying sizes and topologies using different levels of technical accuracy for each conversion technology. In the next section, a Danish case study is presented, illustrating an example model implementation.

\section{Danish Case Study: GreenLab Skive Energy Hub}\label{sec:skive-case-study}
\begin{figure}[ht]
    \centering
    \begin{minipage}{.45\textwidth}
        \centering
        \includegraphics[width=.84\textwidth]{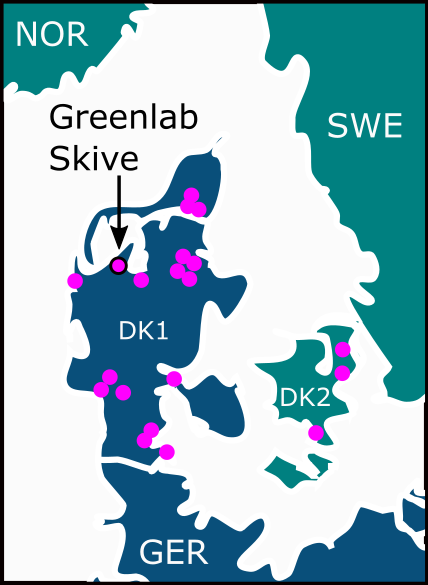} 
        \caption{Location of Greenlab Skive and other P2X sites}
        \label{fig:P2X-location}
    \end{minipage}%
    \hfill
    \begin{minipage}{.52\textwidth}
        \centering
        \includegraphics[width=\textwidth]{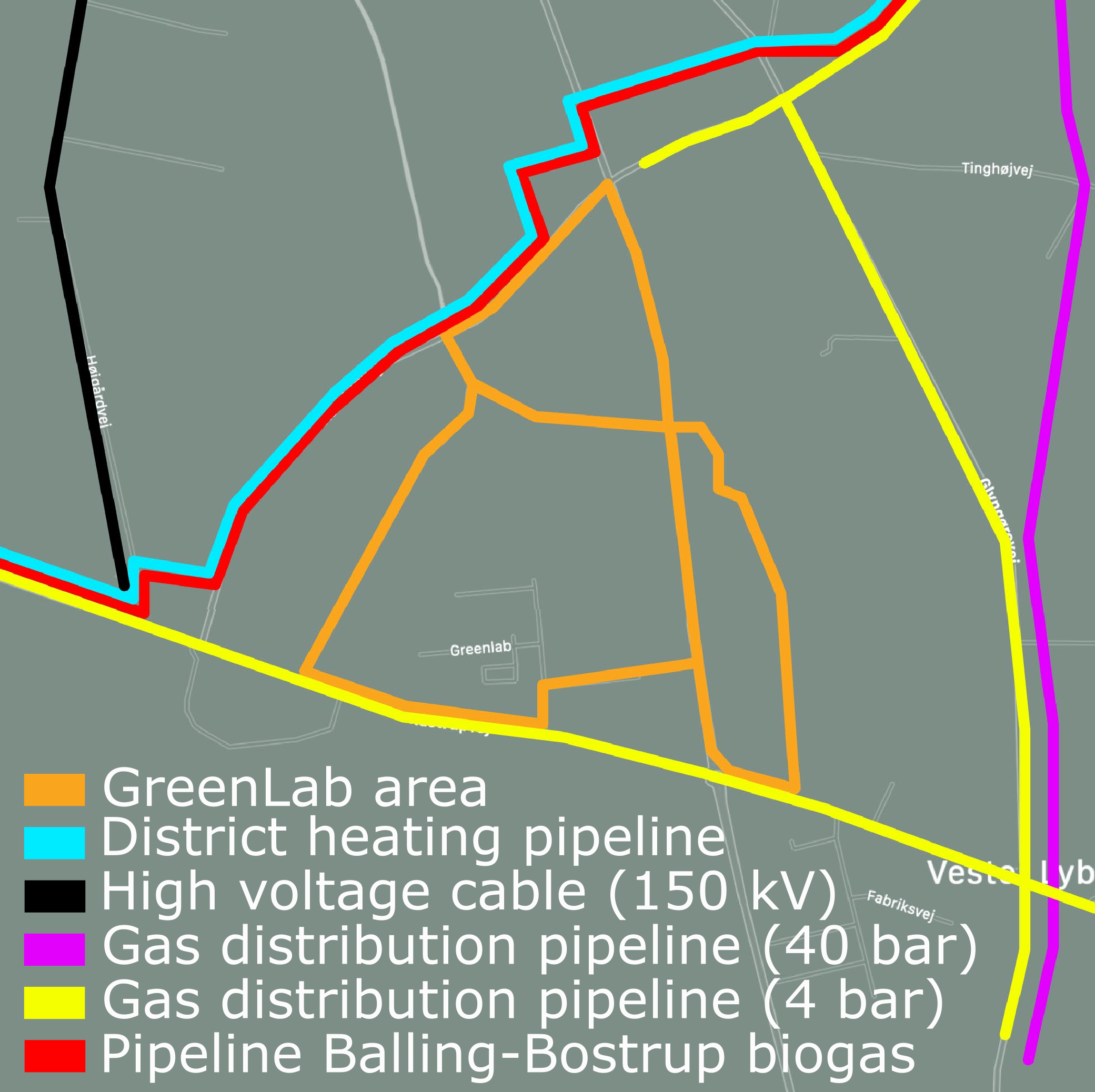} 
        \caption{Greenlab Skive (Future) Infrastructure}
        \label{fig:skive-infrastructure}
    \end{minipage}
\end{figure}

The \citet{danish_government_green_2020} has committed to reduce CO\textsubscript{2} emissions by 70\% by 2030 compared to 1990 levels and to become carbon neutral by 2050. In 2021, its P2X strategy was launched, which outlines at least 4--6 GW of electrolyzer capacity by 2030 \citep{mceu_goverments_2021}. Figure \ref{fig:P2X-location} shows the multitude of pilot projects all over the country\footnote{see \url{https://brintbranchen.dk/danske-brintprojekter/} for updated projects}. 

One of them is GreenLab Skive (GLS), a 60-hectares industrial park in northwestern Denmark. Its goal is to establish an efficient regional \textit{symbiosis network} using RES and P2X technologies. GLS acts as the facilitator and coordinates the exchange of resources within its network of partners (Table~\ref{tab:skive-partners}). Thereupon, bilateral contracts, ensuring a just allocation of resources and benefits, need to be established.

\begin{figure}[ht]
\centering
\includegraphics[scale=0.58]{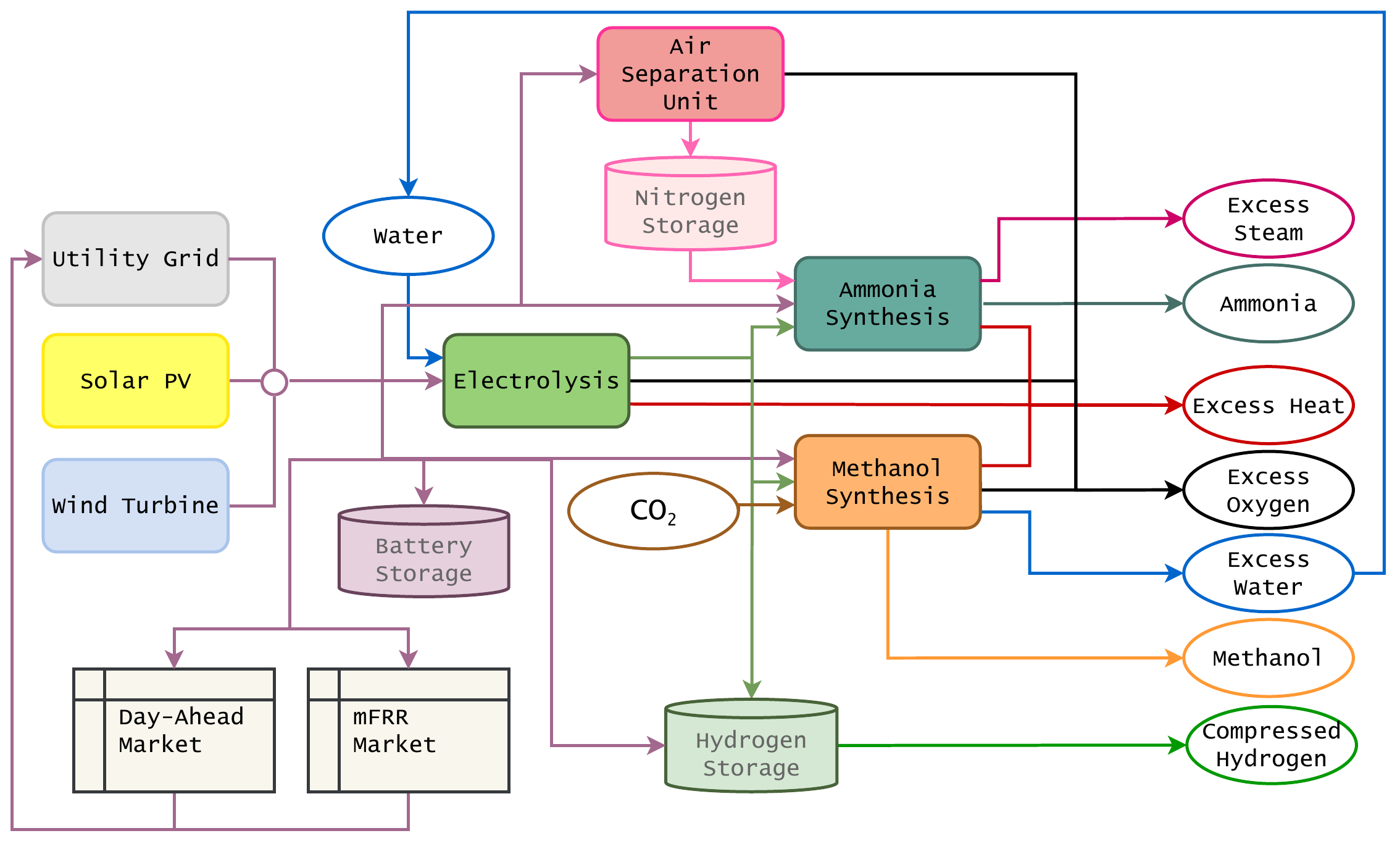} 
\caption{GreenLab Skive Energy Park Technology Flow Chart}
\label{fig:skive-flow-chart}
\end{figure}

Figure \ref{fig:skive-flow-chart} illustrates the current and future technologies of the energy hub, its interlinkages, and the market connections of the hub. In the current system, the following main units are planned: wind and solar energy, electrolysis, methanol synthesis, battery-electric storage, and compressed hydrogen storage. The respective partners, projected capacities, and construction dates can be found in Table~\ref{tab:skive-partners}. Figure  \ref{fig:skive-infrastructure} illustrates its (future) infrastructure connections to the electricity, gas distribution, and district heating networks. 

GLS has five potential revenue streams, it can sell electricity at the day-ahead market or provide ancillary service, it can sell liquefied hydrogen, or use hydrogen to produce and sell ammonia or methanol. The revenues are based on the electricity and feedstock prices of DK 1 in 2019 (Table~\ref{tab:tab:EnergyCarrierPrices}). GLS can source electricity from the utility grid via the day-ahead market. Further capacity expansions and technologies (e.g., ammonia synthesis, carbon capture and storage (CCS)) are planned. In the following, the main conversion technologies are briefly described. 

\begin{table}[ht]
\caption{Skive Energy Hub: Technologies, partners, and current/planned capacities (*approximated based on \citet{klyapovskiy_optimal_2021}}
\label{tab:skive-partners}
\centering
\begin{tabular}{ llll}
\hline
\textbf{Technology} & \textbf{Partners} & \textbf{Capacity} & \textbf{Plan} \\ \hline
Electrolysis & Green Hydrogen Systems, & 12 MW & 2022\\
 & Eurowind \& Lhyfe, & 24 MW & 2023? \\
 &  GreenHyScale Partners & 100 MW &  \\
Methanol Synthesis & European Energy & 10 MW & 2023\\
Ammonia Synthesis & Siemens Gamesa & 20 t/day*\\
Biogas Plant & Eon & 19 mio m\textsuperscript{3} & in use \\
Compr. H2 Storage & Everfuel & 7 tons*\\
Battery Storage & Eurowind & 1.6 MWh\\
Wind Park & Eurowind & 54 MW\\
Solar park & Eurowind & 27 MW\\ \hline
\end{tabular}
\end{table}

\subsection{Electrolyzer} \label{sec:hydrogen}
Hydrogen (H\textsubscript{2}) can be used to produce a number of fuels and other products such as methane, methanol, ammonia, synthetic aviation fuels, and plastics to replace products based on fossil fuels \citep{danish_energy_agency_technology_2022}. It has a high gravimetric energy density of about 120 MJ/kg, which is about three times that of gasoline, but a low volumetric energy density (9 GJ/m\textsuperscript{3} in its liquid state at -253\textdegree C). This is only about one-third that of gasoline \citep{iea_-_advanced_motor_fuels_ammonia_2020}. The global (pure) hydrogen demand in 2020 was estimated at around 90 Mt, mainly used as feedstock in the chemical sector and in the refining industry to produce ammonia or methanol. Today's supply is almost entirely fossil fuel-based \citep{iea_global_2021}.

The most common electrolyzers are alkaline electrolysis cell (AEC) technology, its in-/output characteristics and efficiencies are very similar to Polymer electrolyte membrane electrolysis (PEM), which is the second most common technology \citep{danish_energy_agency_technology_2022}. All electrolyzers use electricity and water as input and have hydrogen, oxygen, and excess heat as output (Table~\ref{tab:skive_electrolyzer}). Since the heat needs to be at a certain temperature to be used in district heating, only a percentage of the heat can be utilized. GLS uses 29x \textit{Green Hydrogen System} A-Series A90 AEC-electrolyzers for the initial 12 MW installation in 2022.

\subsection{Methanol Synthesis} \label{sec:methanol}
Methanol (CH\textsubscript{3}OH or MeOH) could be used straight or blended in fuels, for fuel additives, or for fuel cell applications. It has a lower energy content than hydrogen (20 MJ/kg) but a higher volumetric density (16 GJ/m\textsuperscript{3}) \citep{iea_-_advanced_motor_fuels_methanol_2020}. At ambient temperature, methanol is liquid making it a convenient energy carrier. It can be produced from hydrogen and CO\textsubscript{2} (Table~\ref{tab:skive_methanol}). In 2021, the global production capacity was estimated at around 157 Mt \citep{statista_production_2021}, primarily produced from synthesis gas based on natural gas or coal. 

\subsection{Ammonia Synthesis}
Ammonia (NH\textsubscript{3}) has various applications and could, for example, be used as a carbon-free fuel in maritime transportation \citep{danish_energy_agency_technology_2022}. It has the highest hydrogen content, making it a promising carrier that can be transported in liquid form, it is however highly toxic \citep{incer-valverde_hydrogen-driven_2022}. Its energy content is comparable to methanol (19~MJ/kg), but it has a slightly lower volumetric energy density (13~GJ/m\textsuperscript{3} cooled) \citep{iea_-_advanced_motor_fuels_ammonia_2020}. Since 1913, the Haber-Bosch process is used with hydrogen and nitrogen as input (Table~\ref{tab:skive_ammonia}). In 2021, the global production capacity was estimated at around 236 Mt, mainly used as fertilizer \citep{statista_production_2022}. \\

Next, Section~\ref{sec:numerical-analysis} introduces the numerical analysis based on the Danish case study. Subsequently, Section~\ref{sec:results} discusses the results, limitations, and future research.

\section{Numerical Analysis} \label{sec:numerical-analysis}

\begin{figure}[H]
\centering
\includegraphics[width=\linewidth]{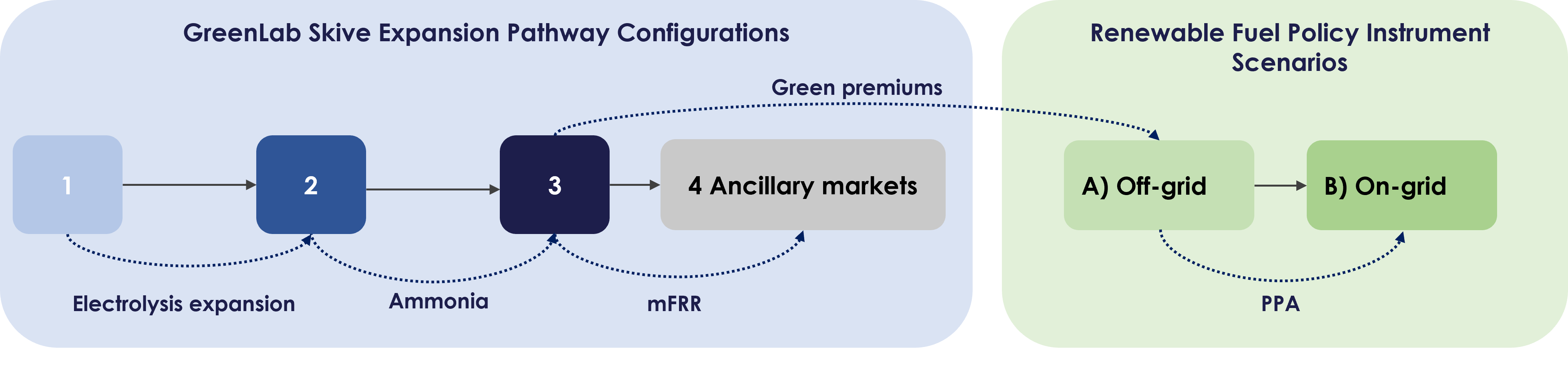}
\caption[GreenLab Skive Energy Flow General Layout]{Graphical description of the four investigated energy hub configurations and two policy scenarios.}
\label{fig:cases}
\end{figure}

In the following, we will introduce the numerical analysis based on the Danish case study introduced in Section~\ref{sec:skive-case-study}. As illustrated in Figure~\ref{fig:cases}, the numerical analysis is split into two parts, first, we investigate four sequential energy hub configurations (Configurations~1 to 4) motivated by the projected expansion pathway of GLS (Table~\ref{tab:skive-partners}). Subsequently, we investigate two different scenarios (Scenario~A and B) of potential policy instruments to support renewable fuel production in energy hubs.


\subsection{GreenLab Skive Expansion Pathway Configurations} \label{sec:case_1}

P2X energy hubs built all over Europe, similar to GreenLab Skive, are expected to demonstrate profitable business cases. Mostly, the hubs start as small pilot sites and scale up by exploiting different revenue streams. Local, national, or European funding oftentimes provides an essential part of the capital expenses. These subsidies can however lead to imbalances when, for example, electrolyzer capacity is expanded, but the RES remain fixed \citep{skov_incentive_2022}. 

To analyze these effects, we consider four hub configurations based on GLS's expansion plans: 1) Initial configuration with 12 MW electrolyzer capacity, 2) Expanded electrolyzer capacity of 100 MW, 3) Expanded electrolyzer capacity plus ammonia synthesis plant, and 4) Expanded version with an ancillary market connection (Figure~\ref{fig:cases}). Configuration~3 provides a holistic picture of the competition between hydrogen and e-fuels production in the energy hub, assuming the respective market prices and technical constraints. We demonstrate the ability of EnerHub2X to handle new network topologies and energy carrier conversion technologies. Configuration~4 illustrates the participation in different electricity markets. For this, we first have to decide which markets are eligible for the GLS energy hub.

In Configurations 1 to 3, the energy hub is only connected to the day-ahead market. The setup of the Danish ancillary services markets for DK~1 (Western Denmark) is described in Table~\ref{tab:dk-energyservices}. In the following, the characteristics of the main products: FCR, aFRR, and mFRR are introduced briefly.

\begin{table}[ht]
\caption{Danish energy system ancillary srvices in DK~1 (2021), announced changes in 2024 in brackets.}
\label{tab:dk-energyservices}
\resizebox{\textwidth}{!}{%
\centering
\begin{tabular}{lccccccc}
\hline
\textbf{Product} & \textbf{up/} & \textbf{Bid} & $\varnothing$-\textbf{Price} & \textbf{Vol.} & \textbf{Ramp} & \textbf{Min.} & \textbf{Min.}\\ 
 & \textbf{down}  & \textbf{length} & [\euro{}/MW/h] & [MW] & [min] & [min] & [MW] \\ \hline
FCR & bi  & 4 h & 27  & $\pm$9 & \cellcolor{red!25}0.5 & 15 & 1 \\
aFRR & bi  & \cellcolor{red!25}1 mth. & 41  & $\pm$100 & 15(5) & 60 & 1 \\
mFRR & up &  1 h & \textbf{5}  & +299 & 15 & 60(15) & 5(1) \\ \hline
\end{tabular}
}
\end{table}

\textbf{Frequency Containment Reserve (FCR)}\footnote{\url{https://energinet.dk/El/Systemydelser/indkob-og-udbud/FCR} (in Danish)} stabilizes the frequency in the electricity grid in case of differences in consumption and production. FCR is activated automatically and can be supplied by both production and consumption units. It is a symmetrical product, which means that both up-and-down-regulation (bi) must be provided. Relatively small amounts are purchased at 8 am the previous day in 4-hour-blocks requiring a fast response time of 30 seconds (Table \ref{tab:dk-energyservices}).

\textbf{Automatic Frequency Restoration Reserve (aFRR)}\footnote{\url{https://energinet.dk/El/Systemydelser/indkob-og-udbud/aFRR} (in Danish)} serves two purposes: 1) Restoring the frequency to 50 Hz and 2) Re-balancing international connections. aFRR is a symmetrical product regulated automatically and can be supplied by both consumption and production units. Medium-sized amounts (bids are normally around 50 MW in DK1) are purchased 1--2~weeks in advance for the whole next month requiring a response time of 15~minutes (Table \ref{tab:dk-energyservices}).

\textbf{Manual Frequency Restoration Reserve (mFRR)}\footnote{\url{https://energinet.dk/El/Systemydelser/indkob-og-udbud/mFRR} (in Danish)} reduces major imbalances and upon activation balances unplanned fluctuations of the electricity system. mFRR is an asynchronous product that is regulated manually and can be delivered by both production and consumption units, and receive a signal upon activation. Energinet only purchases up-regulation capacity in DK 1. Larger-sized amounts are purchased at 9.30 am the previous day requiring a response time of 15 minutes (Table \ref{tab:dk-energyservices}). In both cases, aFRR and mFRR, auctions are capacity payments, and upon activation, separate payments are issued.\\

Overall, we conclude that in the current regulatory setup in Denmark only bidding on the mFRR market would be possible for the energy hub (in addition to the day-ahead market). This is due to the fast ramp-up times needed for FCR services, and due to the requirement of monthly (currently also sometimes a weekly) symmetric bids for the aFRR services. This might change in 2024 when Denmark will (with the other Nordic countries) join ENTSO-E's unification effort for aFRR: PICASSO\footnote{\url{https://www.entsoe.eu/network_codes/eb/picasso/}} and for mFRR: MARI\footnote{\url{https://www.entsoe.eu/network_codes/eb/mari/}} and the bidding windows might become sub-hourly and non-symmetric. The final design is still undecided, however. 

Accordingly, in Configuration 4, EnerHub2X includes the ancillary service market mFRR (see~\ref{sec:appendix:MathematicalModel}). The day ahead market is coupled with the mFRR (reservation and activation), assuming a uniformly distributed activation probability.  

\subsection{Renewable Fuel Policy Instrument Scenarios} \label{sec:case_2}

Analyzing the technical configurations and market participation on the GLS expansion pathway allows us to investigate potential revenue streams and business models. Based on this analysis, in the next step, we investigate which policy instruments could be used to support the production of renewable fuels. An essential part of this analysis is to clearly outline what is considered a \textit{renewable} fuel.

According to the \citet{european_commission_supplementary_2022}, renewable fuels are almost exclusively based on hydrogen produced via electrolysis using renewable electricity. This is the case because the emissions of hydrogen using fossil-based electricity are substantially higher than the emissions of hydrogen produced using natural gas. Therefore, EU regulations on renewable fuels try to prevent the usage of fossil-based electricity by expanding RES or using excess renewable electricity. 

Hence, according to the \citet{european_commission_supplementary_2022} renewable fuels have to comply with three aspects of RES input: \textit{additionality}, \textit{temporal}, and \textit{geographic correlation}. For all of them, the EC suggests (highly controversial) exemptions until the end of 2026. In this study, we focus on the long-term rules. The \textit{additionality} rules make sure that new RES are built instead of re-assigning old installations. It requires either a direct connection (\textit{off-grid}) with RES (no more than 3 years older than the P2X) or using a grid connection (\textit{on-grid}). The latter is valid when the previous year's RES share in the bidding zone was either higher than 90\% or a power purchase agreement (PPA) with RES in the same or adjacent bidding zone without financial support was entered. The rules for \textit{temporal} correlation require that renewable electricity is consumed (either by the electrolyzer or a local storage device) in the same hour, or when the grid price is below 20 EUR per MWh, or during curtailment. Lastly, the \textit{geographic} rules enforce that the electrolyzer is located in the same bidding zone as the RES, an adjacent zone with higher prices, or an adjacent offshore zone. Fulfilling these requirements, GLS would receive guarantees of origin.

In Scenario A, we investigate how green premiums affect the operation and profitability of the energy hub using only its on-site RES (off-grid). In Scenario B, we expand the available renewable electricity to include grid access via a PPA (on-grid). In both cases, a sensitivity analysis of different price premiums for methanol and ammonia is conducted. We assume a competitive PPA price paid on top of the day-ahead price to access renewable electricity.

Next, Section~\ref{sec:results} describes the numerical results and discusses policy implications.


\section{Results and Discussion}\label{sec:results}

In the following, we will analyze the four energy hub configurations, as well as the two policy scenarios and discuss the implications for GLS and P2X energy hubs in general.

\subsection{GreenLab Skive Expansion Pathway Configurations}\label{sec:results-GLS-exp}

In this section, we analyze the energy hub operations taking into account the expansion pathway outlined by GLS. We start with the initial 12 MW electrolyzer capacity (Configuration~1), extend to 100 MW capacity (Configuration~2), and add ammonia production (Configuration~3). Finally, we include the mFRR market (Configuration~4). 

To understand the operations of the energy hub, a detailed analysis of the hub's bottlenecks and potential revenue streams is performed. The competition between revenue streams and energy vectors is characterized by a merit order within the hub. This merit order is, however, dynamic based on variable electricity prices, RES availability, and technical operating constraints.

\begin{figure}[H]
\centering
\includegraphics[width=\linewidth]{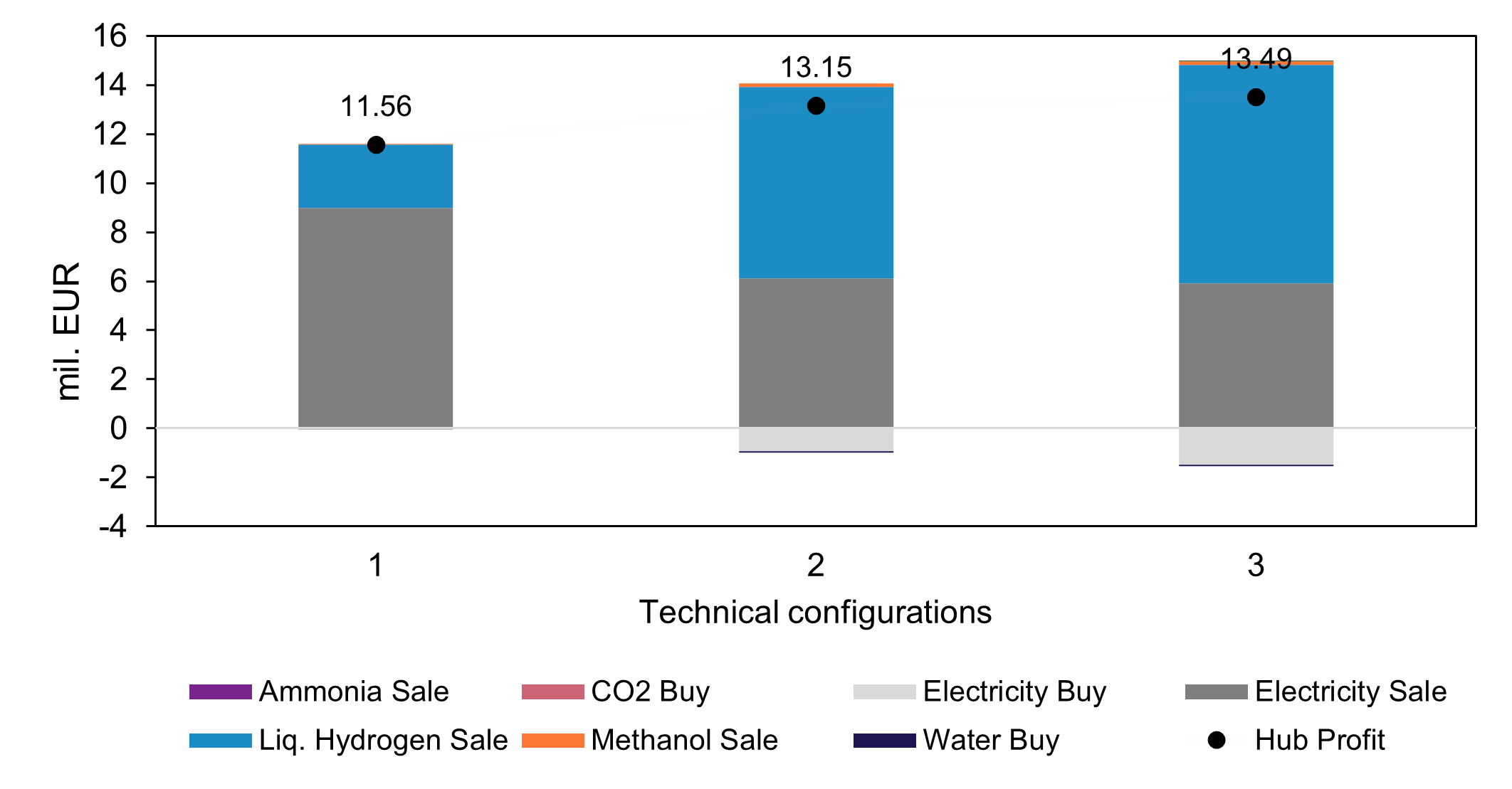}
\caption[GreenLab Skive Energy Flow General Layout]{GLS yearly revenue and cost for the three topology energy hub configurations.}
\label{fig:BaseExpansionScenarios}
\end{figure}

Figure~\ref{fig:BaseExpansionScenarios} shows that although the extended electrolysis capacity (Configurations~2 to 4) allows for more hydrogen and e-fuel production, GLS almost only sells liquefied hydrogen and electricity. 

This is because the profits to be gained selling renewable fuels are relatively low compared to the day-ahead market prices and liquefied hydrogen profits. This way, Figure~\ref{fig:BaseExpansionScenarios} shows that in none of the configurations, a significant amount of e-fuels is produced. Hence, when renewable electricity is available or when electricity prices are low, the hub produces and sells hydrogen. Accordingly, by increasing the electrolyzer capacity in Configurations~2 to 3, a higher share of hydrogen can be produced compared to Configuration~1. In addition, to utilize the higher capacity, more electricity is purchased from the grid at low or negative prices. When electricity prices are high on the other hand, the hub rather sells electricity on the day-ahead market. 

\begin{figure}[H]
\centering
\includegraphics[width=\linewidth]{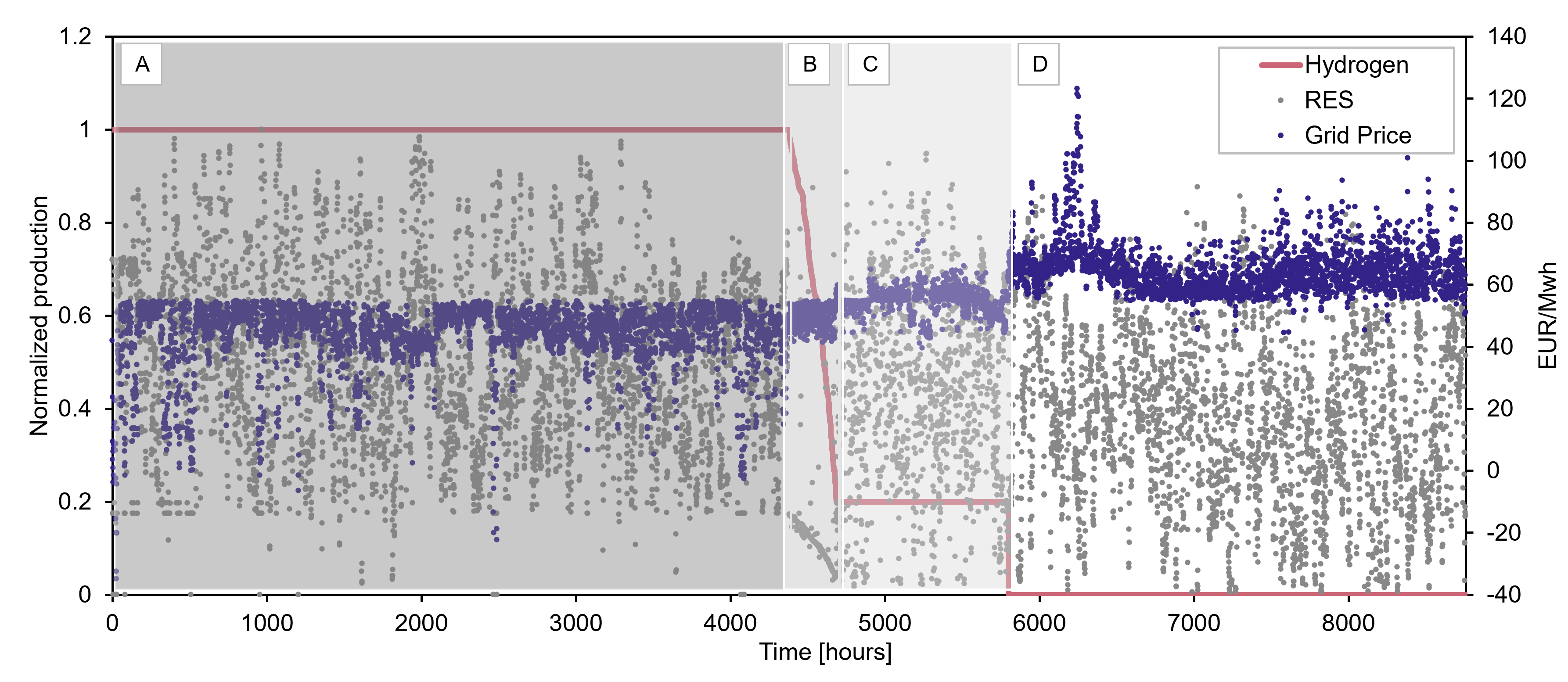}
\caption[GreenLab Skive Energy Flow General Layout]{Operational bottlenecks of Configuration~1, the left y-axis shows the normalized hydrogen and RES production, and the right y-axis shows the grid price (day-ahead price and tariffs).}
\label{fig:BaseOperationAreas}
\end{figure}

In the following, we investigate the bottlenecks of the initial Configuration~1. The electrolyzer acts as intermediate technology between the RES and the P2X units, therefore its operational pattern characterizes the hub's overall behavior. Figure~\ref{fig:BaseOperationAreas} shows the electrolyzer's production duration curve (in descending order from max. to min. utilization) including the electricity prices and RES availability. In Area A (about 50\% of the year), the electrolyzer operates at maximum capacity (i.e., normalized production is equal to 1) with relatively low electricity prices and high RES availability. Area B shows (a small share of hours) with medium electricity prices and decreasing RES availability. Here, the electrolyzer follows the RES availability since purchasing electricity from the grid is too expensive. Area C is characterized by technical limitations and medium to high electricity prices. Here, the hub prefers selling to the grid at high prices but the minimum load requirement forces the electrolyzer to operate on standby to ramp up fast when electricity prices are low. Finally, Area D (about 25\% of the year) shows high electricity prices and no hydrogen production. The electrolyzer is shut down completely and the electricity is sold at the day-ahead market. 

\begin{figure}[H]
\centering
\includegraphics[width=1\linewidth]{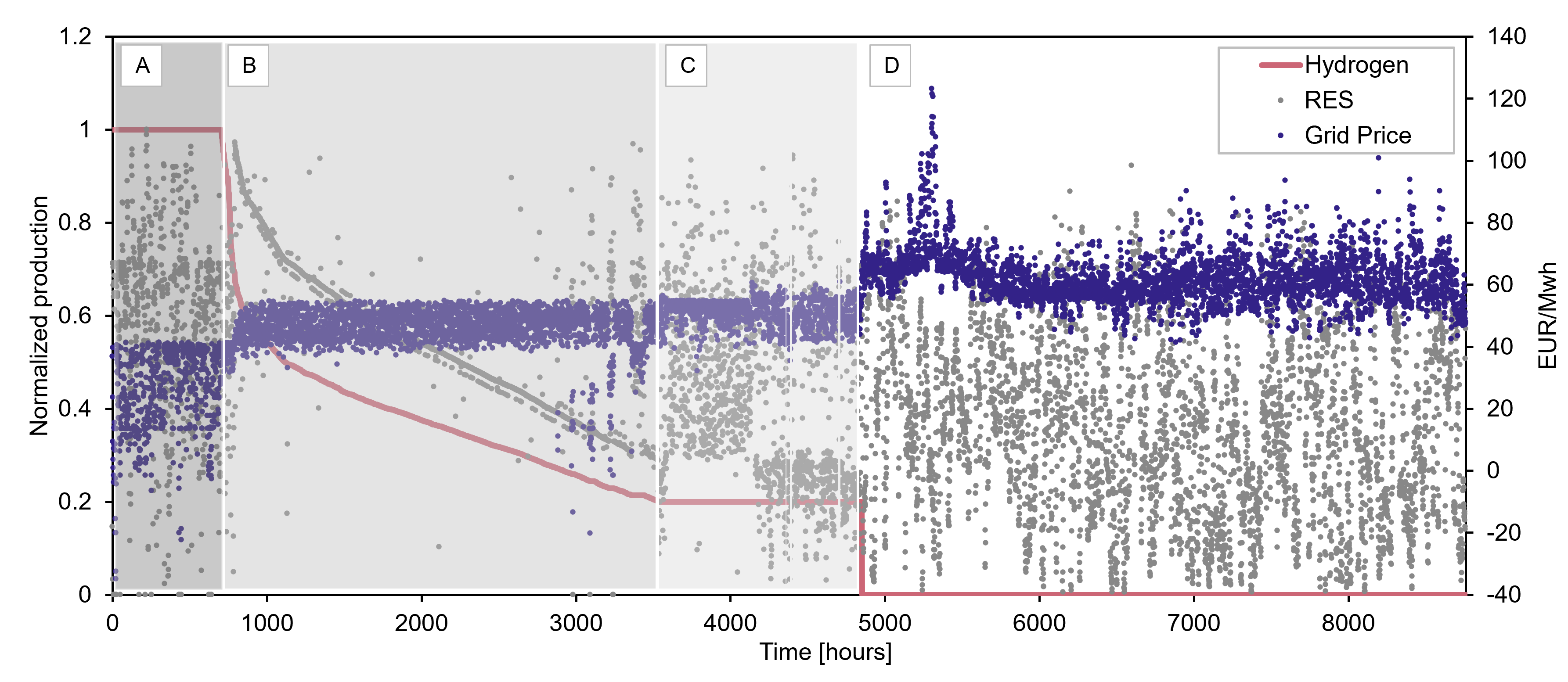}
\caption[GreenLab Skive Energy Flow General Layout]{Operational bottlenecks of Configuration~3.}
\label{fig:Expand+OperationAreas}
\end{figure}

Next, we compare the bottlenecks of Configuration~1 (Figure~\ref{fig:BaseOperationAreas}) with the ones of Configuration~3 (Figure~\ref{fig:Expand+OperationAreas}). Due to the higher electrolyzer capacity, the system is only fully utilized at very low prices (Area A decreases). At medium prices, the hydrogen production decreases with RES availability until the standby load is reached. Area~B is much larger than in Configuration~1 due to the higher capacity. In addition, Area~D is also larger, so the electrolyzer is shut down more often, maybe based on the fact that the minimum load is considerably higher now.

Configuration~3 can also be used to determine the merit order of the renewable fuels in the hub. This merit order is based on the market prices and production costs using a post-processing analysis (Table \ref{tab:EXPAND+Analysis}). In the production hours, the average production costs for ammonia and methanol are 0.24 \euro{}/kg and 0.28\euro{}/kg, respectively.  Production is, however, only profitable in a limited number of hours with low and negative electricity prices (Figure~\ref{fig:DK1pricedurationcurve}). Following the merit order based on the profits of the fuels, hydrogen is dispatched first, then ammonia, and finally methanol (see full-load hours and profits in Table~\ref{tab:EXPAND+Analysis}). 

\begin{table}[H]
\centering
\caption{Market prices, average production cost, full-load hours, total fuels production, and corresponding fuel sales in Configuration~3.}
\label{tab:EXPAND+Analysis}
\resizebox{\textwidth}{!}{%
\begin{tabular}{l p{1.5cm} p{3cm} p{2cm} p{2cm} p{1.5cm}}
\hline
Fuel & Market Price & Average \newline Production Cost & Full-load Hours & Total \newline Production & Fuel Sales \\ 
 &  {[}\euro{}/kg{]} & {[}\euro{}/kg{]} & [h] &  {[}Tons{]}& {[}mil. \euro{}{]} \\ \hline
Liquid Hydrogen & 2.5                & 1.05       & 2045 & 3567 & 8.92\\
Ammonia  & 0.31                    & 0.24       & 223  & 185  & 0.06 \\
Methanol & 0.32                    & 0.28       & 220  & 436 & 0.14  \\ \hline
\end{tabular}
}
\end{table}

To conclude, we find four operational patterns A,B,C, and D in Figures~\ref{fig:BaseOperationAreas} and \ref{fig:Expand+OperationAreas}. These distinct areas should occur in every energy hub but will have different shares based on the hub's configuration. In Area~A, hydrogen production is fully utilized, however, purchasing electricity at low grid prices will lead to non-renewable fuels. Following the recent EC proposal, to produce renewable fuels sourcing from the grid, the grid would have to be 90\% renewable in the previous year or renewable PPAs would have to be concluded. We will evaluate the latter option in Section~\ref{sec:results-policy}. Area~B in Figure~\ref{fig:Expand+OperationAreas} shows that given sufficient capacity, RES availability drives electrolyzer usage. Whereas, Area~C indicates that the technical constraints have a significant influence on operations. Finally, Area~D shows the times when producing renewable fuels including hydrogen is less profitable than selling electricity to the grid. To increase renewable fuels production, this area would have to be converted into a production area. An option to do that would be to increase the profitability of renewable fuels by introducing a premium paid on top of the price for conventional fuels.

We show that the renewable fuels methanol and ammonia are not competitive under current market prices. Without government intervention and support, these fuels will not be produced. To receive green premiums based on guarantees of origin, P2X plants must guarantee that their fuels are renewable. In Section~\ref{sec:results-policy}, we, therefore, apply the latest EC proposal on renewable fuels and analyze the impact of renewable fuel premiums and renewable PPAs on the profitability of the P2X technologies in energy hubs.

First, however, we evaluate if offering ancillary services could be a viable business model for the GLS energy hub.

\subsection{Ancillary Markets Participation} \label{sec:results-ancillary}

To analyze the effects of a fully integrated energy hub, we expand EnerHub2X and allow GLS to offer ancillary services. Based on our reasoning in Section~\ref{sec:case_1}, we focus on the mFRR market. Accordingly, respecting the TSO's current regulations, the P2X assets can only offer up-regulation to the grid. The overall profit gained on the electricity markets consists of three main components (Table~\ref{tab:ResultsAnchillary}): the capacity payments, the day-ahead sales, and the activation payments. The table shows that offering mFRR services results in higher profits. The table also indicates, however, that the capacity and activation payments are small compared to the day-ahead profits because the market volume and prices are much smaller (again this could change with the European market integration).

\begin{table}[H]
\centering
\caption{Results of energy hub participation only on the day-ahead market (Configuration~3), and on the day-ahead and the mFRR market (Configuration~4).}
\label{tab:ResultsAnchillary}
\begin{tabular}{lcc}
\hline
Result                             & Configuration~3 & Configuration~4 \\ \hline
Capacity Payment {[}mil. \euro{}{]}      & -        & 0.11               \\
Day-ahead Profit {[}mil. \euro{}{]}      & 13.49       & 12.24              \\
Activation Payment {[}mil. \euro{}{]}  & -        & 1.44               \\
Overall Market Profit {[}mil. \euro{}{]}          & 13.49       & 13.79              \\ \hline
Total Hydrogen Production {[}tons{]} & 4005        & 4293               \\ \hline
\end{tabular}%
\end{table}

The P2X assets, mainly the electrolysis plant as the most significant electricity consumer, provide demand response to the utility grid. Up-regulation leads to a decrease in the electricity consumption of the P2X assets. Consequently, the P2X assets must operate at least on a minimum level to offer the service and purchase the respective electricity on the day-ahead market. Therefore, the day-ahead profits are lower but compensated for by the additional profits from the mFRR market (Table~\ref{tab:ResultsAnchillary}). Also, the overall hydrogen production is higher, possibly because of not-activated reserves.\\

In this study, the mFRR participation holds minor advantages, however, including the uncertainty in prices and volumes would further diminish them. In addition, one needs to consider that when being activated by the TSO, the hub does not produce fuels. While not being activated, the hub produces fuel, which is, however, not considered renewable due to the use of electricity from the day-ahead market. Therefore, offering up-regulation is not an option for energy hubs that have the goal of being certified as renewable. Only when offering down-regulation (e.g., preventing RES curtailment), the produced fuels would be considered renewable by the EC's proposal \citep{european_commission_supplementary_2022}. Unfortunately, right now the Danish TSO Energinet does not purchase down-regulation.

\subsection{Renewable Fuel Policy Instrument Scenarios} \label{sec:results-policy}

GLS is connected to its own on-site RES, and these will probably be built within three years of the P2X assets and can be used to produce renewable fuels. In Section~\ref{sec:case_1}, we have shown, however, that the production of methanol and ammonia is not economically viable. Therefore, in the following, we will introduce renewable fuel premiums to investigate the effects on the hub's operations only using its own RES (Scenario~A Off-grid).

\begin{figure}[H]
\centering
\includegraphics[width=\linewidth]{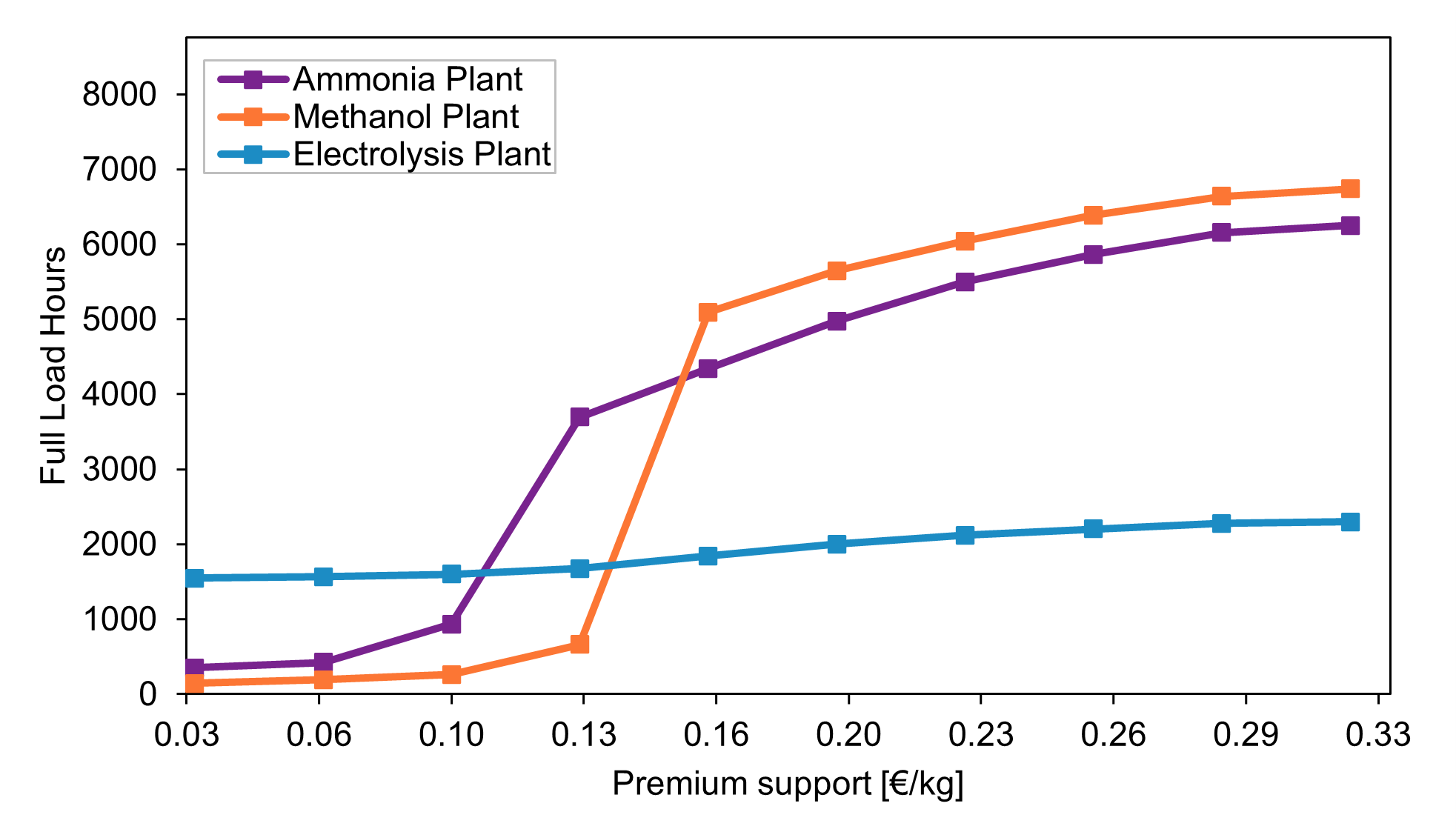}
\caption[GreenLab Skive isolated from the grid vs level of support]{Electrolysis, Ammonia and Methanol plant full load hours assuming different renewable premiums in Scenario A) Off-grid.}
\label{fig:NoGridConntection}
\end{figure}

Figure \ref{fig:NoGridConntection} illustrates the impact of increasing the renewable premiums for ammonia and methanol in 10\%-steps ranging from 10 to 100\%. We analyze the electrolysis and P2X asset's full-load hours (FLH) over the year by dividing the overall production by the daily capacity. We show that a premium of 40\% (0.13 \euro{}/kg) on top of the conventional prices (Table~\ref{tab:EXPAND+Analysis}) significantly increases the production of ammonia that used to be second in the merit order. An increase of 50\% (0.16 \euro{}/kg) leads to an increase in the production of both renewable fuels above 4000 FLH. The graph also shows that the hydrogen FLH are not reduced because hydrogen is still needed to produce the other fuels (Figure~\ref{fig:HydrogenTotalSales}). At this price level, the merit order of the products changes, and selling renewable fuels is preferred over liquid hydrogen. The production of renewable fuels is, however, still restricted by the availability of RES. Therefore, Figure \ref{fig:NoGridConntection} shows a saturation level of FLH even at high premiums.

To increase renewable fuel production, one could increase on-site RES or conclude a renewable PPA (local and without subsidies). We assume the RES capacities are already as large as possible and therefore concentrate on the second option in Scenario~B (On-grid). In this study case, we assume a virtual PPA structure as described in \cite{mendicino_corporate_2019} and \cite{danish_energy_agency_analysis_2019}. The PPA follows the grid price and reflects the cost of the guarantees of origin (GO). Based on \cite{sanchez_potentials_2020} the price of GOs ranged from 0.4--0.5 \euro{}/MWh in 2019. Therefore, we assume GOs cost of 0.5~\euro{}/MWh.

\begin{figure}[H]
\centering
\includegraphics[width=\linewidth]{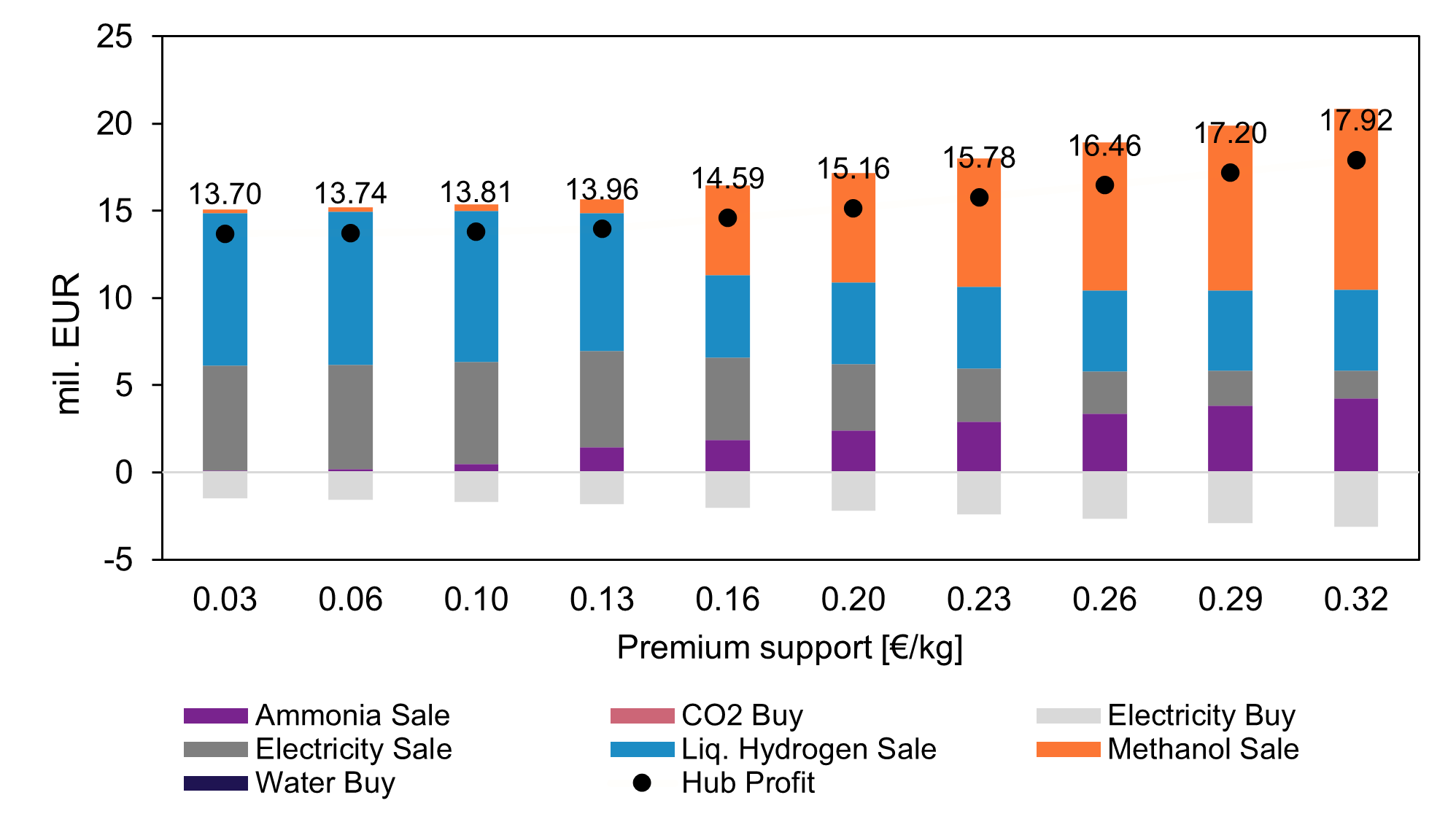}
\caption[GreenLab Skive grid connected with GOs revenues]{Revenues and costs assuming different renewable premiums in Scenario~B) On-grid.}
\label{fig:GOsGridConnectionRevenuevsCosts}
\end{figure}

Figure~\ref{fig:GOsGridConnection} shows that by concluding a PPA the FLH of the system increase because low or negative prices can be utilized. At low premium levels, both scenarios~A and B show similar results. Using a PPA, however, higher premiums eventually lead to full utilization of the supported synthesizers. Figure~\ref{fig:GOsGridConnectionRevenuevsCosts} shows the change in the merit order towards renewable fuels at 0.16 \euro{}/kg and subsequently the continuous reduction of day-ahead market sales.

Overall, we find that renewable premiums increase the production and profitability of renewable fuels in the given GLS energy hub. Using only off-grid RES, their availability sets an upper bound on this expansion. This way, the capacities of P2X assets and on-site RES have to be reconciled carefully. Considering the multitude of influencing factors in these large-scale investments, harmonization might not always be possible. In this case, PPAs could be utilized to balance potential misalignments and allow for higher asset utilization. The EC regulation allows for this option, so the level of renewable premiums can be used to support renewable fuel production. We show that a premium of 0.16 \euro{}/kg, so a price 50\% higher than conventional prices, leads to a substantial increase in renewable fuel production. In the end, however, it is a political decision which level of support should be chosen and which fuels to include in the support scheme. 

\subsection{Limitations and Future Research} \label{sec:reflection}

In this study, GreenLab Skive's operations and future expansion pathway are simulated. EnerHub2X can capture the hub's dynamic evolution by flexibly adjusting its input parameters. Our study does, however, does not consider investments. Oftentimes, the hub's production capacities are already defined and receive substantial subsidies. Therefore, we do not analyze the project's overall economics. Optimal capacity expansion of energy hubs should be considered in future research.

Since the model behavior is highly affected by the price assumptions made regarding the electricity and feedstock prices, an investigation of how well our results transfer to other markets could be interesting.

In addition, the internal and external contracts between consortium members, electricity providers, and consumers could be investigated. The contracts should reflect a just allocation of the hub's profit along the P2X value chain while ensuring the recovery of the individual investments. 

Furthermore, studies could investigate the value of hydrogen and electric battery storage investments and quantify the value of additional flexibility to the energy hub. This could also include more detailed aspects of synergies, for example, considering the integration into a local district heating network.

Finally, future research could focus on the emerging ancillary markets of aFRR: PICASSO and mFRR: MARI. Due to the European integration, the revised structure of the markets and the potentially higher volumes could provide incentives for large consumer units (giga-scale electrolyzers) to provide down-regulation and help tackle system imbalances. Since this might prevent the curtailment of renewable production, the produced renewable fuel would be considered renewable. However, the uncertainty in these markets and the real-time activation magnitude should be properly addressed. 

%

\section{Conclusions and Policy Implications} \label{sec:conclusion}

Recently, the energy landscape is dominated by the discussion on how to rapidly expand renewable hydrogen infrastructure and production to decarbonize the hard-to-abate industry and transport sector. The REPowerEU initiative sets ambitious goals for domestic production and imports, and many EU countries have launched national hydrogen strategies supporting a multitude of pilot to giga-scale projects oftentimes connected to regional energy hubs. However, progress has been slow due to the long-standing ambiguity in the definition of what is considered renewable and the uncertain profitability of renewable fuel production due to new market conditions.

In this study, the multi-commodity flow model, EnerHub2X, simulates the planned expansion pathway of the GreenLab Skive energy hub to gain insights into the profitability of the operation of energy hubs under current market conditions. In addition, this study evaluates different policy instruments to promote renewable fuel production. 

Overall, we find that the electricity prices, renewable energy availability, technical constraints, and feedstock prices strongly influence competition among energy vectors within the hub. The derived merit order of fuels under current market conditions results in a dispatch of hydrogen first, ammonia second, and methanol last. Low renewable energy availability decreases the production of hydrogen substantially and technical constraints can drive system behavior (e.g., forcing standby mode). In addition, a large amount of the renewable electricity produced on-site is not used for renewable fuel production but sold to the grid. Producing electrofuels (ammonia and methanol) is not economically viable under the current market conditions throughout most of the year. We, therefore, conclude that energy hubs must receive renewable fuel premiums to increase fuel production. But in order to do so, the fuels would have to be certified as renewable. A recent proposal by the \citet{european_commission_supplementary_2022} defines the production requirements, taking into account renewable electricity additionality, as well as temporal, and geographic correlation.

Consequently, energy hubs would not be allowed to source electricity from the grid without using their renewable status. Yet, we show that the exclusive use of on-site renewable energy sources (RES) can lead to imbalances and lower utilization of power-to-X (P2X) resources. This can only be prevented by carefully coordinating the capacity expansions of RES and P2X. As this can be challenging due to the multitude of influence factors, we promote using renewable purchase power agreements (PPA) to maintain flexibility. We show that a renewable fuel premium of 0.16~\euro{}/kg corresponding to a 50\% increase in ammonia and methanol prices leads to a substantial increase in renewable fuel production and more than 4000 full-load hours for both fuels. 

In addition, we note that offering ancillary services has only incremental gains. Due to market characteristics, the energy hub in DK~1 could only access the market for manual frequency restoration reserve (mFRR). The Danish TSO Energinet only purchases mFRR up-regulation, which would lead to a decrease in fuel production. In addition, the electricity procured and consumed is not considered renewable. However, the European Commission considers down-regulation as renewable because it prevents renewable energy from being curtailed. This could be an interesting revenue stream in connection with the increase in volume and possibly prices due to the European integration of reserve power markets.

The model's behavior is primarily driven by its price assumptions. The current COVID-19 pandemic and energy crisis and their impact on electricity and fuel prices would have a substantial effect on the results. COVID-19 led to lower electricity prices (2020 DK1: 38 \euro{}/MWh), while the energy crisis causes a historic surge in prices (2021 DK1: 88 \euro{}/MWh)\footnote{\url{https://www.nordpoolgroup.com/en/Market-data1/Dayahead/Area-Prices/ALL1/Yearly/?view=table}}. These fluctuations increase uncertainty even further and highly affect the profitability of individual fuels. On-site RES and long-term PPAs might hedge some of these uncertainties.

Fuel prices for hydrogen, methanol, and ammonia are also affected. The global markets for hydrogen and ammonia are closely linked. The recent rise in natural gas prices leads to higher hydrogen and ammonia prices\footnote{\url{https://www.eia.gov/todayinenergy/detail.php?id=52358}}. However, conventional methanol, a petroleum-based product, has not been as affected \footnote{\url{https://tradingeconomics.com/commodity/methanol}}. This may indicate that hydrogen and ammonia production may already be profitable in the current situation, while methanol production is still dominated by its fossil fuel-based alternative and therefore needs fuel premiums more.

\section*{Data availability}
The code and data that were used for this study are available on GitHub\footnote{\url{https://github.com/IoannisKountouris/}}.

\section*{Declaration of competing interests}
The authors declare that they have no known competing financial interests or personal relationships that could have appeared to influence the work reported in this paper.

\section*{Acknowledgements}
The authors would like to acknowledge financial support from the SuperP2G project that has received funding in the framework of the joint programming initiative ERA-Net Smart Energy Systems’ focus initiative Integrated, Regional Energy Systems, with support from the European Union’s Horizon 2020 research and innovation programme under grant agreement No 775970.


 \bibliographystyle{elsarticle-harv} 
 \bibliography{SuperP2G}

\begin{thebibliography}{61}
\expandafter\ifx\csname natexlab\endcsname\relax\def\natexlab#1{#1}\fi
\providecommand{\url}[1]{\texttt{#1}}
\providecommand{\href}[2]{#2}
\providecommand{\path}[1]{#1}
\providecommand{\DOIprefix}{doi:}
\providecommand{\ArXivprefix}{arXiv:}
\providecommand{\URLprefix}{URL: }
\providecommand{\Pubmedprefix}{pmid:}
\providecommand{\doi}[1]{\href{http://dx.doi.org/#1}{\path{#1}}}
\providecommand{\Pubmed}[1]{\href{pmid:#1}{\path{#1}}}
\providecommand{\bibinfo}[2]{#2}
\ifx\xfnm\relax \def\xfnm[#1]{\unskip,\space#1}\fi
\bibitem[{Aljabery et~al.(2021)Aljabery, Mehrjerdi, Mahdavi and
  Hemmati}]{aljabery_multi_2021}
\bibinfo{author}{Aljabery, A.A.M.}, \bibinfo{author}{Mehrjerdi, H.},
  \bibinfo{author}{Mahdavi, S.}, \bibinfo{author}{Hemmati, R.},
  \bibinfo{year}{2021}.
\newblock \bibinfo{title}{Multi carrier energy systems and energy hubs:
  {Comprehensive} review, survey and recommendations}.
\newblock \bibinfo{journal}{International Journal of Hydrogen Energy}
  \bibinfo{volume}{46}, \bibinfo{pages}{23795--23814}.
\newblock \DOIprefix\doi{10.1016/j.ijhydene.2021.04.178}.
\bibitem[{Almassalkhi and Hiskens(2011)}]{almassalkhi_optimization_2011}
\bibinfo{author}{Almassalkhi, M.}, \bibinfo{author}{Hiskens, I.},
  \bibinfo{year}{2011}.
\newblock \bibinfo{title}{Optimization {Framework} for the {Analysis} of
  {Large}-scale {Networks} of {Energy} {Hubs}}, in:
  \bibinfo{booktitle}{Proceedings: 17th {Power} {Systems} {Computation}
  {Conference} 2011}, \bibinfo{publisher}{Curran Associates},
  \bibinfo{address}{Stockholm}.
\bibitem[{Armijo and Philibert(2020)}]{armijo_flexible_2020}
\bibinfo{author}{Armijo, J.}, \bibinfo{author}{Philibert, C.},
  \bibinfo{year}{2020}.
\newblock \bibinfo{title}{Flexible production of green hydrogen and ammonia
  from variable solar and wind energy: {Case} study of {Chile} and
  {Argentina}}.
\newblock \bibinfo{journal}{International Journal of Hydrogen Energy}
  \bibinfo{volume}{45}, \bibinfo{pages}{1541--1558}.
\newblock \DOIprefix\doi{10.1016/j.ijhydene.2019.11.028}.
\bibitem[{Bellotti et~al.(2022)Bellotti, Rivarolo and
  Magistri}]{bellotti_comparative_2022}
\bibinfo{author}{Bellotti, D.}, \bibinfo{author}{Rivarolo, M.},
  \bibinfo{author}{Magistri, L.}, \bibinfo{year}{2022}.
\newblock \bibinfo{title}{A comparative techno-economic and sensitivity
  analysis of {Power}-to-{X} processes from different energy sources}.
\newblock \bibinfo{journal}{Energy Conversion and Management}
  \bibinfo{volume}{260}, \bibinfo{pages}{115565}.
\newblock \DOIprefix\doi{10.1016/j.enconman.2022.115565}.
\bibitem[{Bramstoft et~al.(2020)Bramstoft, Pizarro-Alonso, Jensen, Ravn and
  Münster}]{bramstoft_modelling_2020}
\bibinfo{author}{Bramstoft, R.}, \bibinfo{author}{Pizarro-Alonso, A.},
  \bibinfo{author}{Jensen, I.G.}, \bibinfo{author}{Ravn, H.},
  \bibinfo{author}{Münster, M.}, \bibinfo{year}{2020}.
\newblock \bibinfo{title}{Modelling of renewable gas and renewable liquid fuels
  in future integrated energy systems}.
\newblock \bibinfo{journal}{Applied Energy} \bibinfo{volume}{268},
  \bibinfo{pages}{114869}.
\newblock \DOIprefix\doi{10.1016/j.apenergy.2020.114869}.
\bibitem[{Chehade et~al.(2019)Chehade, Mansilla, Lucchese, Hilliard and
  Proost}]{chehade_review_2019}
\bibinfo{author}{Chehade, Z.}, \bibinfo{author}{Mansilla, C.},
  \bibinfo{author}{Lucchese, P.}, \bibinfo{author}{Hilliard, S.},
  \bibinfo{author}{Proost, J.}, \bibinfo{year}{2019}.
\newblock \bibinfo{title}{Review and analysis of demonstration projects on
  power-to-{X} pathways in the world}.
\newblock \bibinfo{journal}{International Journal of Hydrogen Energy}
  \bibinfo{volume}{44}, \bibinfo{pages}{27637--27655}.
\newblock \DOIprefix\doi{10.1016/j.ijhydene.2019.08.260}.
\bibitem[{Chen and Yang(2021)}]{chen_power--methanol_2021}
\bibinfo{author}{Chen, C.}, \bibinfo{author}{Yang, A.}, \bibinfo{year}{2021}.
\newblock \bibinfo{title}{Power-to-methanol: {The} role of process flexibility
  in the integration of variable renewable energy into chemical production}.
\newblock \bibinfo{journal}{Energy Conversion and Management}
  \bibinfo{volume}{228}, \bibinfo{pages}{113673}.
\newblock \DOIprefix\doi{10.1016/j.enconman.2020.113673}.
\bibitem[{Chen et~al.(2018)Chen, Wei, Liu, Wu and Mei}]{chen_analyzing_2018}
\bibinfo{author}{Chen, Y.}, \bibinfo{author}{Wei, W.}, \bibinfo{author}{Liu,
  F.}, \bibinfo{author}{Wu, Q.}, \bibinfo{author}{Mei, S.},
  \bibinfo{year}{2018}.
\newblock \bibinfo{title}{Analyzing and validating the economic efficiency of
  managing a cluster of energy hubs in multi-carrier energy systems}.
\newblock \bibinfo{journal}{Applied Energy} \bibinfo{volume}{230},
  \bibinfo{pages}{403--416}.
\newblock \DOIprefix\doi{10.1016/j.apenergy.2018.08.112}.
\bibitem[{{Danish Energy Agency}(2019)}]{danish_energy_agency_analysis_2019}
\bibinfo{author}{{Danish Energy Agency}}, \bibinfo{year}{2019}.
\newblock \bibinfo{title}{Analysis of the {Potential} for {Corporate} {Power}
  {Purchasing} {Agreements} for {Renewable} {Energy} {Production} in
  {Denmark}}.
\newblock \bibinfo{type}{Technical Report}.
\newblock \URLprefix
  \url{https://ens.dk/sites/ens.dk/files/Analyser/corporate_ppa_report_june_2019.pdf}.
\bibitem[{{Danish Energy Agency}(2022)}]{danish_energy_agency_technology_2022}
\bibinfo{author}{{Danish Energy Agency}}, \bibinfo{year}{2022}.
\newblock \bibinfo{title}{Technology {Data} – {Renewable} fuels}.
\newblock \bibinfo{type}{Technical Report}. Danish Energy Agency and Energinet.
\newblock \URLprefix
  \url{https://ens.dk/sites/ens.dk/files/Analyser/technology_data_for_renewable_fuels.pdf}.
  \bibinfo{note}{version 8}.
\bibitem[{{Danish Government}(2020)}]{danish_government_green_2020}
\bibinfo{author}{{Danish Government}}, \bibinfo{year}{2020}.
\newblock \bibinfo{title}{A {Green} and {Sustainable} {World} - {The} {Danish}
  {Government}'s long-term strategy for global climate action}.
\newblock \URLprefix
  \url{https://um.dk/en/-/media/websites/umen/foreign-policy/global-climate-action-strategy/a_green_and_sustainable_world.ashx}.
\bibitem[{Decourt(2019)}]{decourt_weaknesses_2019}
\bibinfo{author}{Decourt, B.}, \bibinfo{year}{2019}.
\newblock \bibinfo{title}{Weaknesses and drivers for power-to-{X} diffusion in
  {Europe}. {Insights} from technological innovation system analysis}.
\newblock \bibinfo{journal}{International Journal of Hydrogen Energy}
  \bibinfo{volume}{44}, \bibinfo{pages}{17411--17430}.
\newblock \DOIprefix\doi{10.1016/j.ijhydene.2019.05.149}.
\bibitem[{Ding et~al.(2022)Ding, Jia, Shahidehpour, Han, Sun and
  Zhang}]{ding_review_2022}
\bibinfo{author}{Ding, T.}, \bibinfo{author}{Jia, W.},
  \bibinfo{author}{Shahidehpour, M.}, \bibinfo{author}{Han, O.},
  \bibinfo{author}{Sun, Y.}, \bibinfo{author}{Zhang, Z.}, \bibinfo{year}{2022}.
\newblock \bibinfo{title}{Review of {Optimization} {Methods} for {Energy} {Hub}
  {Planning}, {Operation}, and {Control}}.
\newblock \bibinfo{journal}{IEEE Transactions on Sustainable Energy} ,
  \bibinfo{pages}{1--1}\DOIprefix\doi{10.1109/TSTE.2022.3172004}.
\bibitem[{ENERGINET(2019)}]{energinet_ptx_2019}
\bibinfo{author}{ENERGINET}, \bibinfo{year}{2019}.
\newblock \bibinfo{title}{{PTX} {IN} {DENMARK} {BEFORE} 2030}.
\newblock \URLprefix
  \url{https://energinet.dk/Om-publikationer/Publikationer/RS-Analyse-April-2019-PtX-i-Danmark-foer-2030}.
\bibitem[{{Energinet}(2019)}]{energinet_winds_2019}
\bibinfo{author}{{Energinet}}, \bibinfo{year}{2019}.
\newblock \bibinfo{title}{Winds of change in a hydrogen perspective - {PtX}
  strategic action plan}.
\newblock \URLprefix
  \url{https://energinet.dk/-/media/4385DB7A333248E7A1A3EA8674E460DA.PDF}.
\bibitem[{{European Commission}(2018)}]{european_commission_directive_2018}
\bibinfo{author}{{European Commission}}, \bibinfo{year}{2018}.
\newblock \bibinfo{title}{Directive ({EU}) 2018/2001 of the {European}
  {Parliament} and of the {Council} of 11 {December} 2018 on the promotion of
  the use of energy from renewable sources}.
\newblock \URLprefix
  \url{https://eur-lex.europa.eu/legal-content/EN/TXT/?uri=uriserv:OJ.L_.2018.328.01.0082.01.ENG}.
\bibitem[{{European Commission}(2022a)}]{european_commission_repowereu_2022}
\bibinfo{author}{{European Commission}}, \bibinfo{year}{2022}a.
\newblock \bibinfo{title}{{REPowerEU} {Plan}}.
\newblock \URLprefix
  \url{https://ec.europa.eu/commission/presscorner/detail/en/IP_22_3131}.
\bibitem[{{European
  Commission}(2022b)}]{european_commission_supplementary_2022}
\bibinfo{author}{{European Commission}}, \bibinfo{year}{2022}b.
\newblock \bibinfo{title}{Supplementary to {Directive} ({EU}) 2018/2001 rules
  on renewable e-fuels}.
\newblock \URLprefix
  \url{https://eur-lex.europa.eu/legal-content/EN/TXT/?uri=PI_COM:Ares(2022)3836651}.
\bibitem[{Gahleitner(2013)}]{gahleitner_hydrogen_2013}
\bibinfo{author}{Gahleitner, G.}, \bibinfo{year}{2013}.
\newblock \bibinfo{title}{Hydrogen from renewable electricity: {An}
  international review of power-to-gas pilot plants for stationary
  applications}.
\newblock \bibinfo{journal}{International Journal of Hydrogen Energy}
  \bibinfo{volume}{38}, \bibinfo{pages}{2039--2061}.
\newblock \DOIprefix\doi{10.1016/j.ijhydene.2012.12.010}.
\bibitem[{Geidl and Andersson(2007)}]{geidl_optimal_2007}
\bibinfo{author}{Geidl, M.}, \bibinfo{author}{Andersson, G.},
  \bibinfo{year}{2007}.
\newblock \bibinfo{title}{Optimal {Power} {Flow} of {Multiple} {Energy}
  {Carriers}}.
\newblock \bibinfo{journal}{IEEE Transactions on Power Systems}
  \bibinfo{volume}{22}, \bibinfo{pages}{145--155}.
\newblock \DOIprefix\doi{10.1109/TPWRS.2006.888988}. \bibinfo{note}{conference
  Name: IEEE Transactions on Power Systems}.
\bibitem[{Ghasemi et~al.(2018)Ghasemi, Banejad and
  Rahimiyan}]{ghasemi_integrated_2018}
\bibinfo{author}{Ghasemi, A.}, \bibinfo{author}{Banejad, M.},
  \bibinfo{author}{Rahimiyan, M.}, \bibinfo{year}{2018}.
\newblock \bibinfo{title}{Integrated energy scheduling under uncertainty in a
  micro energy grid}.
\newblock \bibinfo{journal}{IET Generation, Transmission \& Distribution}
  \bibinfo{volume}{12}, \bibinfo{pages}{2887--2896}.
\newblock \DOIprefix\doi{10.1049/iet-gtd.2017.1631}.
\bibitem[{Glenk and Reichelstein(2019)}]{glenk_economics_2019}
\bibinfo{author}{Glenk, G.}, \bibinfo{author}{Reichelstein, S.},
  \bibinfo{year}{2019}.
\newblock \bibinfo{title}{Economics of converting renewable power to hydrogen}.
\newblock \bibinfo{journal}{Nature Energy} \bibinfo{volume}{4},
  \bibinfo{pages}{216--222}.
\newblock \DOIprefix\doi{10.1038/s41560-019-0326-1}.
\bibitem[{Guelpa et~al.(2019)Guelpa, Bischi, Verda, Chertkov and
  Lund}]{guelpa_towards_2019}
\bibinfo{author}{Guelpa, E.}, \bibinfo{author}{Bischi, A.},
  \bibinfo{author}{Verda, V.}, \bibinfo{author}{Chertkov, M.},
  \bibinfo{author}{Lund, H.}, \bibinfo{year}{2019}.
\newblock \bibinfo{title}{Towards future infrastructures for sustainable
  multi-energy systems: {A} review}.
\newblock \bibinfo{journal}{Energy} \bibinfo{volume}{184},
  \bibinfo{pages}{2--21}.
\newblock \DOIprefix\doi{10.1016/j.energy.2019.05.057}.
\bibitem[{Hajimiragha et~al.(2007)Hajimiragha, Canizares, Fowler, Geidl and
  Andersson}]{hajimiragha_optimal_2007}
\bibinfo{author}{Hajimiragha, A.}, \bibinfo{author}{Canizares, C.},
  \bibinfo{author}{Fowler, M.}, \bibinfo{author}{Geidl, M.},
  \bibinfo{author}{Andersson, G.}, \bibinfo{year}{2007}.
\newblock \bibinfo{title}{Optimal {Energy} {Flow} of integrated energy systems
  with hydrogen economy considerations}, in: \bibinfo{booktitle}{Proceedings:
  2007 {iREP} {Symposium} - {Bulk} {Power} {System} {Dynamics} and {Control} -
  {VII}. {Revitalizing} {Operational} {Reliability}}, pp.
  \bibinfo{pages}{1--11}.
\newblock \DOIprefix\doi{10.1109/IREP.2007.4410517}.
\bibitem[{Hemmes et~al.(2007)Hemmes, Zachariah-Wolf, Geidl and
  Andersson}]{hemmes_towards_2007}
\bibinfo{author}{Hemmes, K.}, \bibinfo{author}{Zachariah-Wolf, J.L.},
  \bibinfo{author}{Geidl, M.}, \bibinfo{author}{Andersson, G.},
  \bibinfo{year}{2007}.
\newblock \bibinfo{title}{Towards multi-source multi-product energy systems}.
\newblock \bibinfo{journal}{International Journal of Hydrogen Energy}
  \bibinfo{volume}{32}, \bibinfo{pages}{1332--1338}.
\newblock \DOIprefix\doi{10.1016/j.ijhydene.2006.10.013}.
\bibitem[{IEA(2019)}]{iea_future_2019}
\bibinfo{author}{IEA}, \bibinfo{year}{2019}.
\newblock \bibinfo{title}{The {Future} of {Hydrogen}}.
\newblock \bibinfo{type}{Technical Report}.
\newblock \URLprefix \url{https://www.iea.org/reports/the-future-of-hydrogen}.
\bibitem[{{IEA}(2021a)}]{iea_global_2021}
\bibinfo{author}{{IEA}}, \bibinfo{year}{2021}a.
\newblock \bibinfo{title}{Global {Hydrogen} {Review}}.
\newblock \URLprefix
  \url{https://iea.blob.core.windows.net/assets/5bd46d7b-906a-4429-abda-e9c507a62341/GlobalHydrogenReview2021.pdf}.
\bibitem[{{IEA}(2021b)}]{iea_hydrogen_2021}
\bibinfo{author}{{IEA}}, \bibinfo{year}{2021}b.
\newblock \bibinfo{title}{Hydrogen {Projects} {Database}}.
\newblock \URLprefix
  \url{https://www.iea.org/data-and-statistics/data-product/hydrogen-projects-database}.
\bibitem[{{IEA}(2021c)}]{iea_net_2021}
\bibinfo{author}{{IEA}}, \bibinfo{year}{2021}c.
\newblock \bibinfo{title}{Net {Zero} by 2050}.
\newblock \bibinfo{type}{Technical Report}. Paris.
\newblock \URLprefix \url{https://www.iea.org/reports/net-zero-by-2050}.
\bibitem[{{IEA - Advanced Motor
  Fuels}(2020a)}]{iea_-_advanced_motor_fuels_ammonia_2020}
\bibinfo{author}{{IEA - Advanced Motor Fuels}}, \bibinfo{year}{2020}a.
\newblock \bibinfo{title}{Ammonia {Application} in {IC} {Engines}}.
\newblock \bibinfo{type}{Technical Report}.
\newblock \URLprefix
  \url{https://iea-amf.org/app/webroot/files/file/other%20publications/Ammonia%20Application%20in%20IC%20Engines.pdf}.
\bibitem[{{IEA - Advanced Motor
  Fuels}(2020b)}]{iea_-_advanced_motor_fuels_methanol_2020}
\bibinfo{author}{{IEA - Advanced Motor Fuels}}, \bibinfo{year}{2020}b.
\newblock \bibinfo{title}{Methanol as {Motor} {Fuel}}.
\newblock \bibinfo{type}{Technical Report}.
\newblock \URLprefix
  \url{https://www.iea-amf.org/app/webroot/files/file/Annex%20Reports/AMF_Annex_56.pdf}.
\bibitem[{Incer-Valverde et~al.(2022)Incer-Valverde, Patiño-Arévalo,
  Tsatsaronis and Morosuk}]{incer-valverde_hydrogen-driven_2022}
\bibinfo{author}{Incer-Valverde, J.}, \bibinfo{author}{Patiño-Arévalo, L.J.},
  \bibinfo{author}{Tsatsaronis, G.}, \bibinfo{author}{Morosuk, T.},
  \bibinfo{year}{2022}.
\newblock \bibinfo{title}{Hydrogen-driven {Power}-to-{X}: {State} of the art
  and multicriteria evaluation of a study case}.
\newblock \bibinfo{journal}{Energy Conversion and Management}
  \bibinfo{volume}{266}, \bibinfo{pages}{115814}.
\newblock \DOIprefix\doi{10.1016/j.enconman.2022.115814}.
\bibitem[{Kauw et~al.(2015)Kauw, Benders and Visser}]{kauw_green_2015}
\bibinfo{author}{Kauw, M.}, \bibinfo{author}{Benders, R.M.J.},
  \bibinfo{author}{Visser, C.}, \bibinfo{year}{2015}.
\newblock \bibinfo{title}{Green methanol from hydrogen and carbon dioxide using
  geothermal energy and/or hydropower in {Iceland} or excess renewable
  electricity in {Germany}}.
\newblock \bibinfo{journal}{Energy} \bibinfo{volume}{90},
  \bibinfo{pages}{208--217}.
\newblock \DOIprefix\doi{10.1016/j.energy.2015.06.002}.
\bibitem[{Klyapovskiy et~al.(2021)Klyapovskiy, Zheng, You and
  Bindner}]{klyapovskiy_optimal_2021}
\bibinfo{author}{Klyapovskiy, S.}, \bibinfo{author}{Zheng, Y.},
  \bibinfo{author}{You, S.}, \bibinfo{author}{Bindner, H.W.},
  \bibinfo{year}{2021}.
\newblock \bibinfo{title}{Optimal operation of the hydrogen-based energy
  management system with {P2X} demand response and ammonia plant}.
\newblock \bibinfo{journal}{Applied Energy} \bibinfo{volume}{304},
  \bibinfo{pages}{117559}.
\newblock \DOIprefix\doi{10.1016/j.apenergy.2021.117559}.
\bibitem[{Krause et~al.(2011)Krause, Andersson, Fröhlich and
  Vaccaro}]{krause_multiple-energy_2011}
\bibinfo{author}{Krause, T.}, \bibinfo{author}{Andersson, G.},
  \bibinfo{author}{Fröhlich, K.}, \bibinfo{author}{Vaccaro, A.},
  \bibinfo{year}{2011}.
\newblock \bibinfo{title}{Multiple-{Energy} {Carriers}: {Modeling} of
  {Production}, {Delivery}, and {Consumption}}.
\newblock \bibinfo{journal}{Proceedings of the IEEE} \bibinfo{volume}{99},
  \bibinfo{pages}{15--27}.
\newblock \DOIprefix\doi{10.1109/JPROC.2010.2083610}.
\bibitem[{Lester et~al.(2020)Lester, Bramstoft and
  Münster}]{lester_analysis_2020}
\bibinfo{author}{Lester, M.S.}, \bibinfo{author}{Bramstoft, R.},
  \bibinfo{author}{Münster, M.}, \bibinfo{year}{2020}.
\newblock \bibinfo{title}{Analysis on {Electrofuels} in {Future} {Energy}
  {Systems}: {A} 2050 {Case} {Study}}.
\newblock \bibinfo{journal}{Energy} \bibinfo{volume}{199},
  \bibinfo{pages}{117408}.
\newblock \DOIprefix\doi{10.1016/j.energy.2020.117408}.
\bibitem[{Lin et~al.(2021)Lin, Wu, Chen, Yang, Guo, Lv, Lu, Song and
  McElroy}]{lin_economic_2021}
\bibinfo{author}{Lin, H.}, \bibinfo{author}{Wu, Q.}, \bibinfo{author}{Chen,
  X.}, \bibinfo{author}{Yang, X.}, \bibinfo{author}{Guo, X.},
  \bibinfo{author}{Lv, J.}, \bibinfo{author}{Lu, T.}, \bibinfo{author}{Song,
  S.}, \bibinfo{author}{McElroy, M.}, \bibinfo{year}{2021}.
\newblock \bibinfo{title}{Economic and technological feasibility of using
  power-to-hydrogen technology under higher wind penetration in {China}}.
\newblock \bibinfo{journal}{Renewable Energy} \bibinfo{volume}{173},
  \bibinfo{pages}{569--580}.
\newblock \DOIprefix\doi{10.1016/j.renene.2021.04.015}.
\bibitem[{Lück et~al.(2017)Lück, Larscheid, Maaz and
  Moser}]{luck_economic_2017}
\bibinfo{author}{Lück, L.}, \bibinfo{author}{Larscheid, P.},
  \bibinfo{author}{Maaz, A.}, \bibinfo{author}{Moser, A.},
  \bibinfo{year}{2017}.
\newblock \bibinfo{title}{Economic potential of water electrolysis within
  future electricity markets}, in: \bibinfo{booktitle}{2017 14th
  {International} {Conference} on the {European} {Energy} {Market} ({EEM})},
  pp. \bibinfo{pages}{1--6}.
\newblock \DOIprefix\doi{10.1109/EEM.2017.7981950}. \bibinfo{note}{iSSN:
  2165-4093}.
\bibitem[{Mancarella(2014)}]{mancarella_mes_2014}
\bibinfo{author}{Mancarella, P.}, \bibinfo{year}{2014}.
\newblock \bibinfo{title}{{MES} (multi-energy systems): {An} overview of
  concepts and evaluation models}.
\newblock \bibinfo{journal}{Energy} \bibinfo{volume}{65},
  \bibinfo{pages}{1--17}.
\newblock \DOIprefix\doi{10.1016/j.energy.2013.10.041}.
\bibitem[{Maniyali et~al.(2013)Maniyali, Almansoori, Fowler and
  Elkamel}]{maniyali_energy_2013}
\bibinfo{author}{Maniyali, Y.}, \bibinfo{author}{Almansoori, A.},
  \bibinfo{author}{Fowler, M.}, \bibinfo{author}{Elkamel, A.},
  \bibinfo{year}{2013}.
\newblock \bibinfo{title}{Energy {Hub} {Based} on {Nuclear} {Energy} and
  {Hydrogen} {Energy} {Storage}}.
\newblock \bibinfo{journal}{Industrial \& Engineering Chemistry Research}
  \bibinfo{volume}{52}, \bibinfo{pages}{7470--7481}.
\newblock \DOIprefix\doi{10.1021/ie302161n}.
\bibitem[{Maroufmashat et~al.(2016)Maroufmashat, Fowler, Sattari~Khavas,
  Elkamel, Roshandel and Hajimiragha}]{maroufmashat_mixed_2016}
\bibinfo{author}{Maroufmashat, A.}, \bibinfo{author}{Fowler, M.},
  \bibinfo{author}{Sattari~Khavas, S.}, \bibinfo{author}{Elkamel, A.},
  \bibinfo{author}{Roshandel, R.}, \bibinfo{author}{Hajimiragha, A.},
  \bibinfo{year}{2016}.
\newblock \bibinfo{title}{Mixed integer linear programing based approach for
  optimal planning and operation of a smart urban energy network to support
  the hydrogen economy}.
\newblock \bibinfo{journal}{International Journal of Hydrogen Energy}
  \bibinfo{volume}{41}, \bibinfo{pages}{7700--7716}.
\newblock \DOIprefix\doi{10.1016/j.ijhydene.2015.08.038}.
\bibitem[{Mendicino et~al.(2019)Mendicino, Menniti, Pinnarelli and
  Sorrentino}]{mendicino_corporate_2019}
\bibinfo{author}{Mendicino, L.}, \bibinfo{author}{Menniti, D.},
  \bibinfo{author}{Pinnarelli, A.}, \bibinfo{author}{Sorrentino, N.},
  \bibinfo{year}{2019}.
\newblock \bibinfo{title}{Corporate power purchase agreement: {Formulation} of
  the related levelized cost of energy and its application to a real life case
  study}.
\newblock \bibinfo{journal}{Applied Energy} \bibinfo{volume}{253},
  \bibinfo{pages}{113577}.
\newblock \DOIprefix\doi{10.1016/j.apenergy.2019.113577}.
\bibitem[{Mohammadi et~al.(2017)Mohammadi, Noorollahi, Mohammadi-ivatloo and
  Yousefi}]{mohammadi_energy_2017}
\bibinfo{author}{Mohammadi, M.}, \bibinfo{author}{Noorollahi, Y.},
  \bibinfo{author}{Mohammadi-ivatloo, B.}, \bibinfo{author}{Yousefi, H.},
  \bibinfo{year}{2017}.
\newblock \bibinfo{title}{Energy hub: {From} a model to a concept – {A}
  review}.
\newblock \bibinfo{journal}{Renewable and Sustainable Energy Reviews}
  \bibinfo{volume}{80}, \bibinfo{pages}{1512--1527}.
\newblock \DOIprefix\doi{10.1016/j.rser.2017.07.030}.
\bibitem[{Morgan et~al.(2017)Morgan, Manwell and
  McGowan}]{morgan_sustainable_2017}
\bibinfo{author}{Morgan, E.R.}, \bibinfo{author}{Manwell, J.F.},
  \bibinfo{author}{McGowan, J.G.}, \bibinfo{year}{2017}.
\newblock \bibinfo{title}{Sustainable {Ammonia} {Production} from {U}.{S}.
  {Offshore} {Wind} {Farms}: {A} {Techno}-{Economic} {Review}}.
\newblock \bibinfo{journal}{ACS Sustainable Chemistry \& Engineering}
  \bibinfo{volume}{5}, \bibinfo{pages}{9554--9567}.
\newblock \DOIprefix\doi{10.1021/acssuschemeng.7b02070}.
\bibitem[{Nasir et~al.(2022)Nasir, Rezaee~Jordehi, Matin, Tabar, Tostado-Véliz
  and Mansouri}]{nasir_optimal_2022}
\bibinfo{author}{Nasir, M.}, \bibinfo{author}{Rezaee~Jordehi, A.},
  \bibinfo{author}{Matin, S.A.A.}, \bibinfo{author}{Tabar, V.S.},
  \bibinfo{author}{Tostado-Véliz, M.}, \bibinfo{author}{Mansouri, S.A.},
  \bibinfo{year}{2022}.
\newblock \bibinfo{title}{Optimal operation of energy hubs including parking
  lots for hydrogen vehicles and responsive demands}.
\newblock \bibinfo{journal}{Journal of Energy Storage} \bibinfo{volume}{50},
  \bibinfo{pages}{104630}.
\newblock \DOIprefix\doi{10.1016/j.est.2022.104630}.
\bibitem[{Ordóñez et~al.(2021)Ordóñez, Halfdanarson, Ganzer,
  Guillén-Gosálbez, Dowell and Shah}]{ordonez_carbon_2021}
\bibinfo{author}{Ordóñez, D.F.}, \bibinfo{author}{Halfdanarson, T.},
  \bibinfo{author}{Ganzer, C.}, \bibinfo{author}{Guillén-Gosálbez, G.},
  \bibinfo{author}{Dowell, N.M.}, \bibinfo{author}{Shah, N.},
  \bibinfo{year}{2021}.
\newblock \bibinfo{title}{Carbon or {Nitrogen}-based e-fuels? {A} comparative
  techno-economic and full environmental assessment}, in:
  \bibinfo{editor}{Türkay, M.}, \bibinfo{editor}{Gani, R.} (Eds.),
  \bibinfo{booktitle}{Computer {Aided} {Chemical} {Engineering}}.
  \bibinfo{publisher}{Elsevier}. volume~\bibinfo{volume}{50} of
  \textit{\bibinfo{series}{31 {European} {Symposium} on {Computer} {Aided}
  {Process} {Engineering}}}, pp. \bibinfo{pages}{1623--1628}.
\newblock \DOIprefix\doi{10.1016/B978-0-323-88506-5.50251-5}.
\bibitem[{Osman et~al.(2020)Osman, Sgouridis and
  Sleptchenko}]{osman_scaling_2020}
\bibinfo{author}{Osman, O.}, \bibinfo{author}{Sgouridis, S.},
  \bibinfo{author}{Sleptchenko, A.}, \bibinfo{year}{2020}.
\newblock \bibinfo{title}{Scaling the production of renewable ammonia: {A}
  techno-economic optimization applied in regions with high insolation}.
\newblock \bibinfo{journal}{Journal of Cleaner Production}
  \bibinfo{volume}{271}, \bibinfo{pages}{121627}.
\newblock \DOIprefix\doi{10.1016/j.jclepro.2020.121627}.
\bibitem[{Parra et~al.(2017)Parra, Zhang, Bauer and
  Patel}]{parra_integrated_2017}
\bibinfo{author}{Parra, D.}, \bibinfo{author}{Zhang, X.},
  \bibinfo{author}{Bauer, C.}, \bibinfo{author}{Patel, M.K.},
  \bibinfo{year}{2017}.
\newblock \bibinfo{title}{An integrated techno-economic and life cycle
  environmental assessment of power-to-gas systems}.
\newblock \bibinfo{journal}{Applied Energy} \bibinfo{volume}{193},
  \bibinfo{pages}{440--454}.
\newblock \DOIprefix\doi{10.1016/j.apenergy.2017.02.063}.
\bibitem[{Pérez-Fortes et~al.(2016)Pérez-Fortes, Schöneberger, Boulamanti
  and Tzimas}]{perez-fortes_methanol_2016}
\bibinfo{author}{Pérez-Fortes, M.}, \bibinfo{author}{Schöneberger, J.C.},
  \bibinfo{author}{Boulamanti, A.}, \bibinfo{author}{Tzimas, E.},
  \bibinfo{year}{2016}.
\newblock \bibinfo{title}{Methanol synthesis using captured {CO2} as raw
  material: {Techno}-economic and environmental assessment}.
\newblock \bibinfo{journal}{Applied Energy} \bibinfo{volume}{161},
  \bibinfo{pages}{718--732}.
\newblock \DOIprefix\doi{10.1016/j.apenergy.2015.07.067}.
\bibitem[{Sadeghi et~al.(2019)Sadeghi, Rashidinejad, Moeini-Aghtaie and
  Abdollahi}]{sadeghi_energy_2019}
\bibinfo{author}{Sadeghi, H.}, \bibinfo{author}{Rashidinejad, M.},
  \bibinfo{author}{Moeini-Aghtaie, M.}, \bibinfo{author}{Abdollahi, A.},
  \bibinfo{year}{2019}.
\newblock \bibinfo{title}{The energy hub: {An} extensive survey on the
  state-of-the-art}.
\newblock \bibinfo{journal}{Applied Thermal Engineering} \bibinfo{volume}{161},
  \bibinfo{pages}{114071}.
\newblock \DOIprefix\doi{10.1016/j.applthermaleng.2019.114071}.
\bibitem[{Salmon and Bañares-Alcántara(2021)}]{salmon_green_2021}
\bibinfo{author}{Salmon, N.}, \bibinfo{author}{Bañares-Alcántara, R.},
  \bibinfo{year}{2021}.
\newblock \bibinfo{title}{Green ammonia as a spatial energy vector: a review}.
\newblock \bibinfo{journal}{Sustainable Energy \& Fuels} \bibinfo{volume}{5},
  \bibinfo{pages}{2814--2839}.
\newblock \DOIprefix\doi{10.1039/D1SE00345C}.
\bibitem[{Scolaro and Kittner(2022)}]{scolaro_optimizing_2022}
\bibinfo{author}{Scolaro, M.}, \bibinfo{author}{Kittner, N.},
  \bibinfo{year}{2022}.
\newblock \bibinfo{title}{Optimizing hybrid offshore wind farms for
  cost-competitive hydrogen production in {Germany}}.
\newblock \bibinfo{journal}{International Journal of Hydrogen Energy}
  \bibinfo{volume}{47}, \bibinfo{pages}{6478--6493}.
\newblock \DOIprefix\doi{10.1016/j.ijhydene.2021.12.062}.
\bibitem[{Sharif et~al.(2014)Sharif, Almansoori, Fowler, Elkamel and
  Alrafea}]{sharif_design_2014}
\bibinfo{author}{Sharif, A.}, \bibinfo{author}{Almansoori, A.},
  \bibinfo{author}{Fowler, M.}, \bibinfo{author}{Elkamel, A.},
  \bibinfo{author}{Alrafea, K.}, \bibinfo{year}{2014}.
\newblock \bibinfo{title}{Design of an energy hub based on natural gas and
  renewable energy sources}.
\newblock \bibinfo{journal}{International Journal of Energy Research}
  \bibinfo{volume}{38}, \bibinfo{pages}{363--373}.
\newblock \DOIprefix\doi{10.1002/er.3050}.
\bibitem[{Skov and Schneider(2022)}]{skov_incentive_2022}
\bibinfo{author}{Skov, I.R.}, \bibinfo{author}{Schneider, N.},
  \bibinfo{year}{2022}.
\newblock \bibinfo{title}{Incentive structures for power-to-{X} and e-fuel
  pathways for transport in {EU} and member states}.
\newblock \bibinfo{journal}{Energy Policy} \bibinfo{volume}{168},
  \bibinfo{pages}{113121}.
\newblock \DOIprefix\doi{10.1016/j.enpol.2022.113121}.
\bibitem[{Son et~al.(2021)Son, Oh, Acquah, Fan, Kim and Kim}]{son_multi_2021}
\bibinfo{author}{Son, Y.G.}, \bibinfo{author}{Oh, B.C.},
  \bibinfo{author}{Acquah, M.A.}, \bibinfo{author}{Fan, R.},
  \bibinfo{author}{Kim, D.M.}, \bibinfo{author}{Kim, S.Y.},
  \bibinfo{year}{2021}.
\newblock \bibinfo{title}{Multi {Energy} {System} {With} an {Associated}
  {Energy} {Hub}: {A} {Review}}.
\newblock \bibinfo{journal}{IEEE Access} \bibinfo{volume}{9},
  \bibinfo{pages}{127753--127766}.
\newblock \DOIprefix\doi{10.1109/ACCESS.2021.3108142}.
\bibitem[{{Statista}(2021)}]{statista_production_2021}
\bibinfo{author}{{Statista}}, \bibinfo{year}{2021}.
\newblock \bibinfo{title}{Production capacity of methanol worldwide from 2018
  to 2020, with a forecast for 2030}.
\newblock \URLprefix
  \url{https://www.statista.com/statistics/1065891/global-methanol-production-capacity/}.
\bibitem[{{Statista}(2022)}]{statista_production_2022}
\bibinfo{author}{{Statista}}, \bibinfo{year}{2022}.
\newblock \bibinfo{title}{Production capacity of ammonia worldwide from 2018 to
  2021, with a forecast for 2026 and 2030}.
\newblock \URLprefix
  \url{https://www.statista.com/statistics/1065865/ammonia-production-capacity-globally/}.
\bibitem[{Sánchez and Jacobsen(2020)}]{sanchez_potentials_2020}
\bibinfo{author}{Sánchez, I.N.B.}, \bibinfo{author}{Jacobsen, H.K.},
  \bibinfo{year}{2020}.
\newblock \bibinfo{title}{Potentials for transborder green gas and hydrogen
  certificate markets}.
\newblock \bibinfo{type}{Technical Report}.
\bibitem[{Zhang et~al.(2020)Zhang, Wang, Van~herle, Maréchal and
  Desideri}]{zhang_techno-economic_2020}
\bibinfo{author}{Zhang, H.}, \bibinfo{author}{Wang, L.},
  \bibinfo{author}{Van~herle, J.}, \bibinfo{author}{Maréchal, F.},
  \bibinfo{author}{Desideri, U.}, \bibinfo{year}{2020}.
\newblock \bibinfo{title}{Techno-economic comparison of green ammonia
  production processes}.
\newblock \bibinfo{journal}{Applied Energy} \bibinfo{volume}{259},
  \bibinfo{pages}{114135}.
\newblock \DOIprefix\doi{10.1016/j.apenergy.2019.114135}.
\bibitem[{Zhang et~al.(2019)Zhang, Martín and
  Grossmann}]{zhang_integrated_2019}
\bibinfo{author}{Zhang, Q.}, \bibinfo{author}{Martín, M.},
  \bibinfo{author}{Grossmann, I.E.}, \bibinfo{year}{2019}.
\newblock \bibinfo{title}{Integrated design and operation of renewables-based
  fuels and power production networks}.
\newblock \bibinfo{journal}{Computers \& Chemical Engineering}
  \bibinfo{volume}{122}, \bibinfo{pages}{80--92}.
\newblock \DOIprefix\doi{10.1016/j.compchemeng.2018.06.018}.
\bibitem[{Zheng et~al.(2022)Zheng, You, Li, Bindner and
  Münster}]{zheng_data-driven_2022}
\bibinfo{author}{Zheng, Y.}, \bibinfo{author}{You, S.}, \bibinfo{author}{Li,
  X.}, \bibinfo{author}{Bindner, H.W.}, \bibinfo{author}{Münster, M.},
  \bibinfo{year}{2022}.
\newblock \bibinfo{title}{Data-driven robust optimization for optimal
  scheduling of power to methanol}.
\newblock \bibinfo{journal}{Energy Conversion and Management}
  \bibinfo{volume}{256}, \bibinfo{pages}{115338}.
\newblock \DOIprefix\doi{10.1016/j.enconman.2022.115338}.

\end{thebibliography}

\appendix

\section{Additional Figures}\label{sec:appendix:figures}

\begin{figure}[H]
\centering
\includegraphics[width=\linewidth]{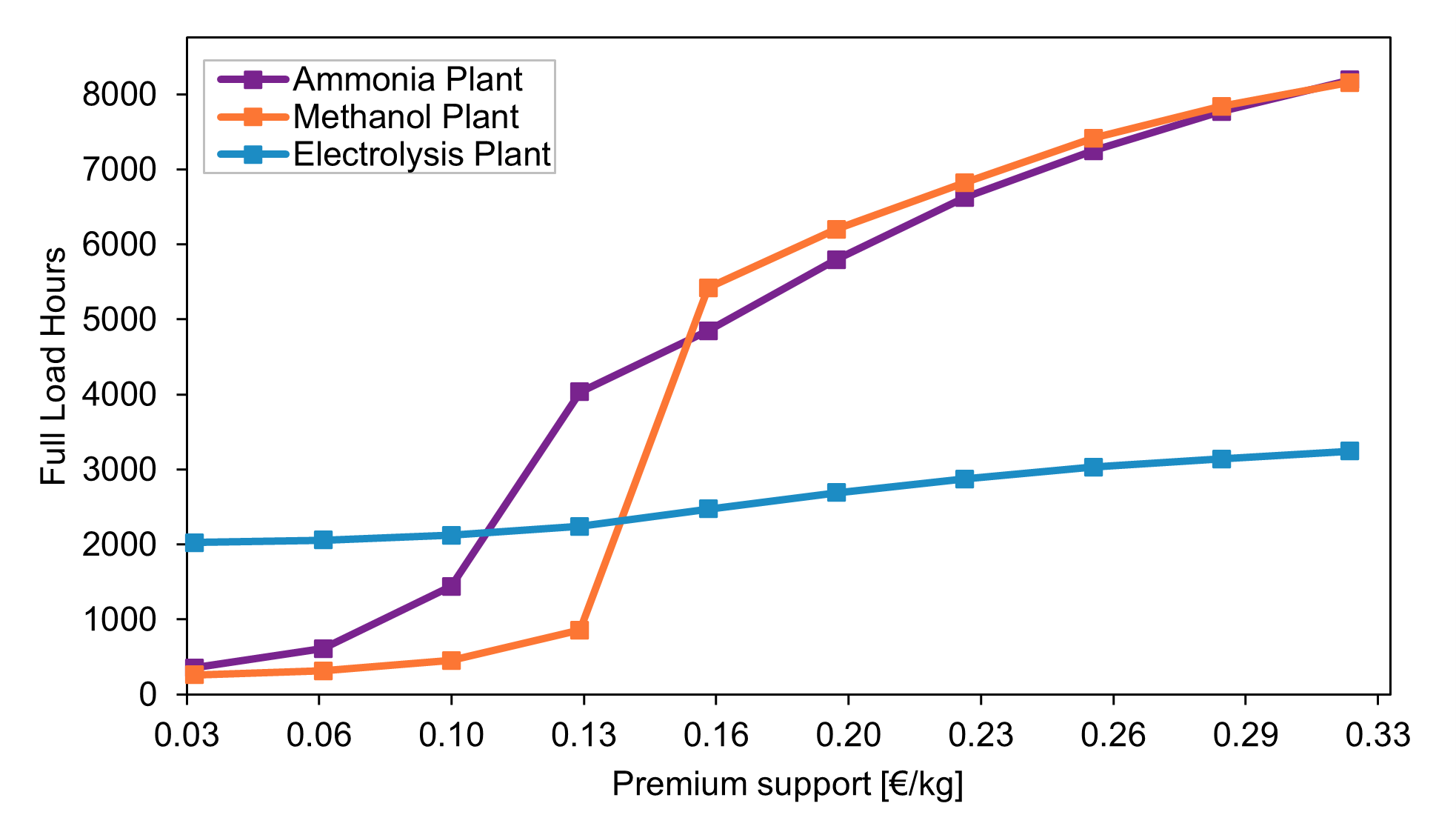}
\caption[GreenLab Skive grid connected with GOs vs level of support]{Case 3b - Offsite: Ammonia and methanol plants full load hours (y axis) vs level of support (x axis)}
\label{fig:GOsGridConnection}
\end{figure}

\begin{figure}[H]
\centering
\includegraphics[width=0.9\linewidth]{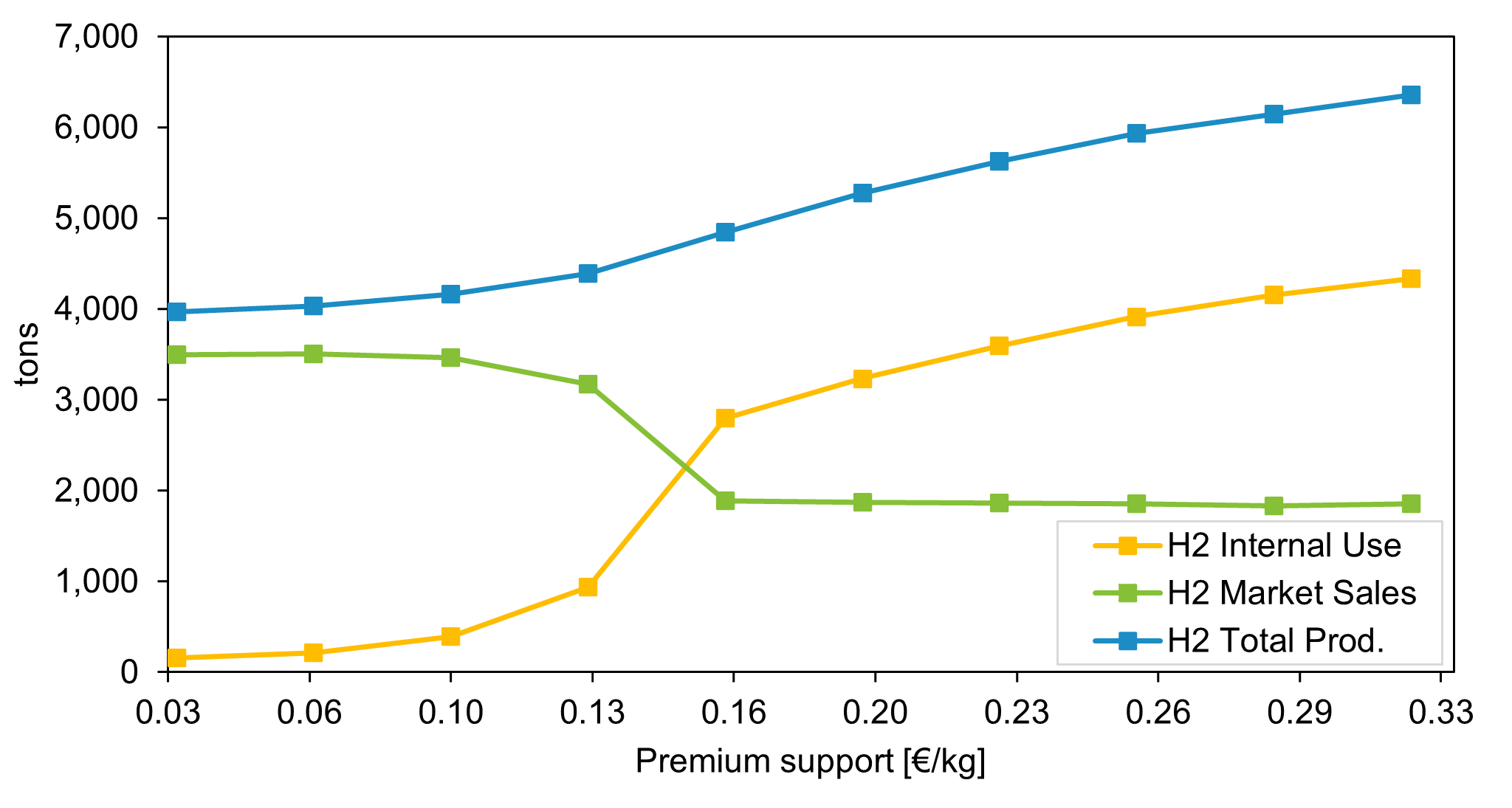}
\caption[GreenLab Skive grid connected with GOs vs level of support]{Case 3b - Offsite: Net yearly Hydrogen production composed by internal consumption and external sales}
\label{fig:HydrogenTotalSales}
\end{figure}

\begin{figure}[H]
\centering
\includegraphics[width=1\linewidth]{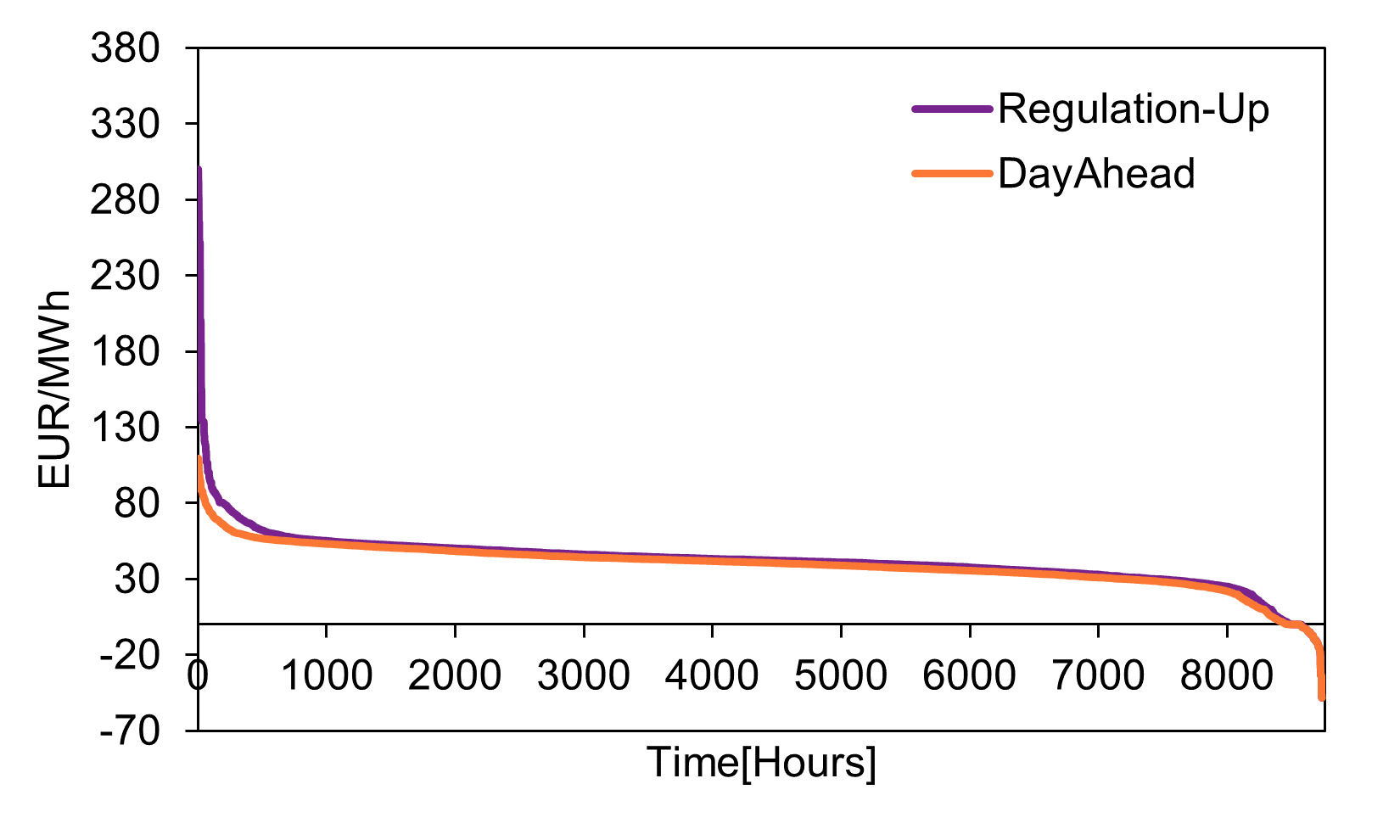}
\caption[Day-ahead price]{Day-ahead market and positive regulation Up electricity price duration curves, 2019 DK1}
\label{fig:DK1pricedurationcurve}
\end{figure}

\section{Future hydrogen related projects}\label{sec:IEAdata}

\clearpage \newpage
\begin{table}[h]
\caption{Hydrogen project expansion stages over 10 MW\textsubscript{e} capacity and starting production until 2025, underlined are already operational (CIP...Copenhagen Infrastructure Partners)}
\label{tab:h2-projects}
\centering
\begin{minipage}{\linewidth}
\resizebox{\textwidth}{!}{%
\begin{tabular}{|lllccll|}
\hline
\textbf{Project} & \textbf{Lead role/ } & \textbf{Location} & \textbf{Size} & \textbf{Type} & \textbf{Output} & \textbf{Year}\\ 
 & \textbf{Tech. provider} & & \textbf{ [MW\textsubscript{e}]} & &  & \\ 
\hline
\multicolumn{7}{|l|}{\textbf{Mobility}}\\
\hline \hline
\href{https://www.global.toshiba/ww/news/energy/2020/03/news-20200307-01.html}{H2 Energy Research Field}   & Toshiba & Fukushima (JPN)  & 10 & ALK & H\textsubscript{2} & \underline{2020}\\
\hline
\href{https://www.airliquide.com/group/press-releases-news/2021-01-26/air-liquide-inaugurates-worlds-largest-low-carbon-hydrogen-membrane-based-production-unit-canada}{Air Liquide Bécancour} & Air Liquide/ Cummins (US) & Bécancour (CAN)  & 20 & PEM & H\textsubscript{2} & \underline{2021}\\
\hline
Green fuels for DK & Ørsted \& mobility companies/ & Copenhagen (DNK) & 10 & ? & H\textsubscript{2}&2023\\
 & ? &  & 250 &&e-fuels& 2027\\ 
 &  & & 1300 &&e-fuels& 2030\\ 
\hline
\multicolumn{7}{|l|}{\textbf{Ports}}\\
\hline \hline
? & H2 Energy (CHE)/ Plug (USA) & Esbjerg (DNK) & 1000 & PEM & H\textsubscript{2} & 2024\\
\hline
HØST PtX Esbjerg & CIP/  Plug (USA) & Esbjerg (DNK) & 1000 & PEM & NH\textsubscript{3} & 2025\\
\hline
\multicolumn{7}{|l|}{\textbf{Refurbished refineries}}\\
\hline \hline
\href{https://refhyne.eu/}{Refhyne} & Shell/ ITM (GBR) & Wesseling (DEU) & 10 & PEM & H\textsubscript{2} & \underline{2021}\\
 & &  & 100 &&& 2025\\ 
\hline
\href{https://www.everfuel.com/projects/hysynergy/}{HySynergy}& Everfuel, Crossbridge Energy/ & Fredericia (DNK) & 20 & PEM & H\textsubscript{2} & 2022\\
&  ? &  & 300 &?&H\textsubscript{2} & 2025 \\
&  && 1000 &?&H\textsubscript{2} & 2030 \\ 
\hline
\href{https://www.h2bulletin.com/rwe-receives-funding-approval-for-the-lingen-facility/}{GET H2 Lingen} & RWE/ Sunfire (ALK),  & Lingen (DEU) & 10 & ALK & H\textsubscript{2} & 2023\\
 & Linde (PEM)  &  & 4 & PEM & H\textsubscript{2}& 2023\\ 
&  && 300 &?& &2026 \\ 
\hline
\href{https://www.westkueste100.de/en/}{Westk\"uste 100} & Shell/ EDF (FRA) & Heide (DEU) & 30 & ALK & H\textsubscript{2}& 2023\\
& &  & 700 &&H\textsubscript{2}& 2030\\ 
\hline
\href{https://lingen-green-hydrogen.com/}{Lingen Green Hydrogen}  & BP, Ørsted (DNK)/ & Lingen (DEU) & 50 & ? & H\textsubscript{2}& 2024\\
 & ?  & & 100 &&H\textsubscript{2}& 2025\\ 
&  && 500 &&H\textsubscript{2}& 2027 \\ 
\hline
\multicolumn{7}{|l|}{\textbf{Ammonia production}}\\
\hline \hline
Cachimayo Plant & Industrias Cachimayo & Cusco (PER) & 25 & ALK & NH\textsubscript{3} & \underline{1965}\\
\hline
\href{https://www.iberdrola.com/about-us/lines-business/flagship-projects/puertollano-green-hydrogen-plant}{Green hydrogen plant} & Iberdrola/ Nel ASA (NO) & Puertollano (ES) & 20 & PEM & NH\textsubscript{3} & \underline{2022}\\
\hline
HyEx & Engie (FRA), Enaex/ & Antofagasta (CHL) & 26 & ? &NH\textsubscript{3}& 2025\\
& ? &  & 1600 && NH\textsubscript{3}&2030\\ 
\hline
\end{tabular}
}
\end{minipage}
\end{table}
\clearpage

\section{Data and GreenLab Skive technical parameters }\label{sec:appendix:skive-data}

\begin{table}[H]
\centering
\caption{Energy carriers prices for 2019}
\label{tab:tab:EnergyCarrierPrices}
\resizebox{\textwidth}{!}{%
\begin{tabular}{clcc}
\hline
Energy Carrier & Units & Market Price 2019 & Source \\ \hline
Day-ahead electricity, DK1               &  \euro{}/MWh     & Dynamic              & (\tablefootnote{see \url{https://www.energidataservice.dk/tso-electricity/Elspotprices/} })        \\
Electricity grid tarrifs                      & \euro{}/MWh  &  13.5                &  \cite{energinet_ptx_2019}\\
Frequency Restoration Reserves (mFRR), DK1    & \euro{}/MW   & Dynamic             &  (\tablefootnote{see \url{https://www.energidataservice.dk/tso-electricity/MfrrReservesDK1/} })\\
Regulating prices                             & \euro{}/MWh  & Dynamic             &  (\tablefootnote{ \url{https://www.energidataservice.dk/tso-electricity/RealtimeMarket/} })\\                      
Hydrogen                                      & \euro{}/kg   & 2.5                 & \cite{iea_future_2019}   \\
Ammonia                                       & \euro{}/kg   & 0.31                &  (\tablefootnote{ \url{https://www.icis.com/explore/resources/news/2019/09/20/10420486/tfi-19-global-ammonia-prices-stable-to-firm-ahead-of-first-q4-spot-business/} })\\
Methanol                                      & \euro{}/kg   & 0.32                &  (\tablefootnote{see \url{https://www.methanol.org/methanol-price-supply-demand/} })    \\ 
Water                                         & \euro{}/kg   & 0.078               & \cite{glenk_economics_2019} \\
CO2                                           & \euro{}/kg   & 0.27                & Internal contract price \\
 \hline
\end{tabular}%
}
\end{table}

\begin{table}[H]
\caption{Mass and energy flows of the electrolyzer based on \cite{danish_energy_agency_technology_2022}}
\label{tab:skive_electrolyzer}
\centering
\begin{tabular}{ll|cc|cc}
\hline
&& \multicolumn{2}{c}{\textbf{Production}} & \multicolumn{2}{c}{\textbf{Flow balance}} \\ 
\textbf{Carrier} & \textbf{Units} & IN & OUT & IN & OUT  \\ \hline
Electricity & MWh & 53.06 & & 85\% \\
Water & Tons & 9.55 & & 15\% \\ 
\hline
Hydrogen & Tons &  & \textbf{1} & & 5\% \\
Heat & MWh & & 9.76 & & 44\% \\
Oxygen & Tons & & 11.26 & & 51\% \\
\hline
 & &  &  & 100\%  & 100\% \\
\end{tabular}
\end{table}

\begin{table}[H]
\caption{Mass and energy flows of the methanol synthesis based on \cite{perez-fortes_methanol_2016} analysis}
\label{tab:skive_methanol}
\centering
\begin{tabular}{ll|cc|cc}
\hline
&& \multicolumn{2}{c}{\textbf{Production}} & \multicolumn{2}{c}{\textbf{Flow balance}} \\ 
\textbf{Carrier} & \textbf{Units} & IN & OUT & IN & OUT  \\ \hline
Electricity & MWh & 0.169 & & 10\%  \\
Hydrogen & Tons & 0.192 & & 11\% \\
CO\textsubscript{2} & Tons & 1.37 & & 79\% \\
\hline
Methanol & Tons & & \textbf{1} && 64\% \\
Water & Tons & & 0.569 & & 36\% \\ 
\hline
 & &  &  & 100\%  & 100\% \\
\end{tabular}
\end{table}

\begin{table}[H]
\caption{Mass and energy flows of the ammonia synthesis according to \cite{danish_energy_agency_technology_2022}}
\label{tab:skive_ammonia}
\centering
\begin{tabular}{ll|cc|cc}
\hline
&& \multicolumn{2}{c}{\textbf{Production}} & \multicolumn{2}{c}{\textbf{Flow balance}} \\ 
\textbf{Carrier} & \textbf{Units} & IN & OUT & IN & OUT  \\ \hline
Electricity & MWh & 0.34 & & 25\%  \\
Hydrogen & Tons & 0.18 & & 13\% \\
Nitrogen & Tons & 0.84 & & 62\% \\
\hline
Ammonia & Tons & & \textbf{1} && 52\% \\
Steam & MWh & & 0.69 & & 36\% \\ 
Heat & MWh & & 0.24 & & 12\% \\
\hline
 & &  &  & 100\%  & 100\% \\
\end{tabular}
\end{table}

\clearpage \newpage
\begin{sidewaystable}
\begin{table}[H]
\centering
\caption{GreenLab Skive technology parameters}
\label{tab:GreenLabTechnicalParameters}
\resizebox{\textwidth}{!}{%
\begin{tabular}{cccccccccc}
\hline
Parameter & Units & WindTurbine & SolarPV & Electrolysis & MethanolSynthesis & AmmoniaSynthesis & AirSeparationUnit & HydrogenStorage & ElectricStorage \\ \hline
Capacity          & MW or tons/h & 54.60 & 27.00 & 0.24  & 1.98  & 0.83  & 1.00 & 7.00  & 1.60 \\
Minimum Operation & \%load/h     & -     & -     & 20.00 & -     & 25.00 & -    & -     & -    \\
RampRate          & \%load/h     & -     & -     & 95.00 & -     & 20.00 & -    & -     & -    \\
CapacityUP        & \%load/h     & -     & -     & 20.00 & 20.00 & 20.00 & -    & 20.00 & -    \\
CapacityDown      & \%load/h     & -     & -     & 20.00 & 20.00 & 20.00 & -    & 20.00 & -    \\
StorageCap        & MWh or tons  & -     & -     & -     & -     & -     & -    & 7.00  & 1.60 \\ \hline
\end{tabular}%
}
\end{table}
\end{sidewaystable}
\clearpage

\section{EnerHub2X Mathematical Formulation}\label{sec:appendix:MathematicalModel}


\nomenclature[A]{\(g\)}{Available technologies/Process}
\nomenclature[A]{\(e\)}{Energy carriers/Fuels}
\nomenclature[A]{\(t\)}{Dispatching periods}
\nomenclature[A]{\(a\)}{Areas}
\nomenclature[A]{\(d\)}{Direction of flows, \{input/output\}}

\nomenclature[B]{\(\mathbb{G}\)}{Set of dispatchable technology units}

\nomenclature[B]{\(\mathbb{G}^{R}\)}{Subset of dispatchable  technology units with Ramping technical constraint}

\nomenclature[B]{\(\mathbb{G}^{S}\)}{Subset of storage technology units}

\nomenclature[B]{\(\mathbb{G}^{UC}\)}{Subset of dispatchable technology units with unit commitment constraints (on/off)}

\nomenclature[B]{\(\mathbb{G}^{LD}\)}{Subset of dispatchable  technology units with efficiency varying based on load dependency}

\nomenclature[B]{\(\mathbb{G}^{AS^{+}}\)}{Subset of dispatchable  technology units able to provide Up regulation ancillary services}

\nomenclature[B]{\(\mathbb{G}^{AS^{-}}\)}{Subset of dispatchable  technology units able to provide Down regulationancillary services}

\nomenclature[B]{\(\mathbb{G}^{Var}\)}{Subset of variable energy technology units}

\nomenclature[B]{\(\mathbb{G}_{a}\)}{Mapping set of technologies to areas}

\nomenclature[B]{\(\mathbb{E}\)}{Set of fuels}

\nomenclature[B]{\(\mathbb{E}^{In}_{g}\)}{Mapping set of feedstock fuels to corresponding technologies}

\nomenclature[B]{\(\mathbb{E}^{Out}_{g}\)}{Mapping set of produced fuels from technologies}

\nomenclature[B]{\(\mathbb{E}^{Buy}_{a}\)}{Mapping set of feedstock fuels purchased from areas}

\nomenclature[B]{\(\mathbb{E}^{Sell}_{a}\)}{Mapping set of end product fuels sold to area}

\nomenclature[B]{\(\mathbb{E}^{D}\)}{Subset of fuels associated with demand constraint}

\nomenclature[B]{\(\mathbb{E}^{Var}\)}{Subset of fuels characterized by variable production}

\nomenclature[B]{\(\mathbb{E}^{Def}_{g}\)}{Mapping set of fuels defining the technical characteristics of technology}

\nomenclature[B]{\(\mathbb{F}_{(a,a)}\)}{Mapping set of fuels flow from area to area}

\nomenclature[B]{\(\mathbb{K}\)}{Piecewise linear approximation brakepoints of the load dependent production efficiency curve}

\nomenclature[C]{\(\sigma^{in}_{g,e}\)}{Percentage of fuel $e $ used as a feedstock in technology $g$}

\nomenclature[C]{\(\sigma^{out}_{g,e}\)}{Percentage of fuel $e $ produced by technology $g$}

\nomenclature[C]{\(F^{out}_{g,e}\)}{Fixed amount of fuel $e$ produced by technology $g$}

\nomenclature[C]{\(F^{in}_{g,e}\)}{Fixed amount of fuel $e $ used as a feedstock in technology $g$}

\nomenclature[C]{\(\Theta_{g}\)}{Percentage of activity state change from consumption to production of technology $g$}

\nomenclature[C]{\(U_{g,k}\)}{Load levels of technology $g$ at break point $k$}

\nomenclature[C]{\(\Theta_{g,k}\)}{Percentage of activity state change from consumption to production of technology $g$ in breakpoint $k$}

\nomenclature[C]{\(P^{max}_{g}\)}{Maximum production level/capacity of technology $g$}

\nomenclature[C]{\(R^{+}_{g}/R^{-}_{g}\)}{Maximum up/down reserve capacity of technology $g$}


\nomenclature[C]{\(P^{min}_{g}\)}{Minimum production level of technology $g$}

\nomenclature[C]{\(SOC^{max}_{g}\)}{Maximum volume of storage technology $g$}

\nomenclature[C]{\(SOC^{init}_{g}\)}{Initial volume of storage technology $g$ for $t=1$}

\nomenclature[C]{\(R^{up}_{g}/R^{down}_{g}\)}{Ramp up/down rate limit of technology $g$}


\nomenclature[C]{\(C^{var}_{g}\)}{Variable operating cost of $g $}

\nomenclature[C]{\(C^{start}_{g}\)}{Start-up cost of dispatchable technology $g$}


\nomenclature[C]{\(\pi^{buy}_{a,e,t}\)}{Price of feedstock fuel $e$ bought from area $a$ in period $t$}

\nomenclature[C]{\(\pi^{sell}_{a,e,t}\)}{Price of end product fuel $e$ sold at area $a$ in period $t$}

\nomenclature[C]{\(D_{a,e,t}\)}{Demand at area $a$  of end product fuel $e$ in period $t$}

\nomenclature[C]{\(X^{cap}_{a,e,t}\)}{Maximum flow of fuel $e$ from area $a$ in period $t$}

\nomenclature[C]{\(\pi^{activation}_{t}\)}{Price for reserve activation in period $t$}

\nomenclature[C]{\(\pi^{+}_{t}/\pi^{-}_{t}\)}{mFRR Up/Down regulation reserve capacity payment in period $t$}


\nomenclature[C]{\(\pi^{\omega}_{t}\)}{Activation probability in period $t$}

\nomenclature[C]{\(P^{profile}_{g,t}\)}{Capacity factor of technology $g$ in period $t$}

\nomenclature[D]{\(x^{in}_{g,e,t}\)}{Amount of feedstock fuel $e $ consumed in technology $g$ in period $t$}

\nomenclature[D]{\(x^{out}_{g,e,t}\)}{Amount of produced fuel $e $ from technology $g $ in period $t$}

\nomenclature[D]{\(x^{total}_{g,t}\)}{Level of activity of technology $g$ in period $t$}

\nomenclature[D]{\(soc_{g,t}\)}{State of charge for technology $g$ in period $t$}

\nomenclature[D]{\(f^{flow}_{a,a',e,t}\)}{Flow from area $a$ to $a'$ of fuel $e$ in period $t$}

\nomenclature[D]{\(q^{buy}_{a,e,t}\)}{Quantity of feedstock fuel $e \in \mathbb{E}^{buy}_{a}$ bought from area $a$ in period $t$}

\nomenclature[D]{\(q^{sell}_{a,e,t}\)}{Quantity of end product fuel $e \in \mathbb{E}^{sell}_{a}$ sold to area $a$ in period $t$}

\nomenclature[D]{\(c^{start}_{g,t}\)}{Start up cost of technology $g$ in period $t$}

\nomenclature[D]{\(o_{g,t}\)}{Binary indicating online/offline status of technology $g$ in period $t$}
\nomenclature[D]{\(u_{g,t}\)}{Binary indicating when status up occurs of technology $g$ in period $t$}

\nomenclature[D]{\(w^{sos2}_{g,t,k}\)}{Activity level of technology $g$ in period $t$ at break point $k$}

\nomenclature[D]{\(\nu_{g,t}\)}{Binary indicating charging or discharging status of storage technology $g$ in period $t$}

\nomenclature[D]{\(x^{in^{+}}_{g,e,t}/x^{in^{-}}_{g,e,t}\)}{Amount of feedstock fuel $e$ consumed in technology $g$ in period $t$ for activation of Up/Down regulation service}

\nomenclature[D]{\(x^{out^{+}}_{g,e,t}/x^{out^{-}}_{g,e,t}\)}{Amount of produced fuel $e$ from technology $g$ in period $t$  for activation of Up/Down-regulation service}

\nomenclature[D]{\(x^{total^{+}}_{g,t}/x^{total^{-}}_{g,t}\)}{Level of activity of technology $g$ in period $t$  for activation of Up/Down-regulation service}

\nomenclature[D]{\(x^{cap^{+}}_{g,t}/x^{cap^{-}}_{g,t}\)}{Capacity reserved $g $ in period $t$ for providing Up/Down-regulation service}

\nomenclature[D]{\(q^{{buy}^{+}}_{a,e,t}/q^{{buy}^{-}}_{a,e,t}\)}{Quantity of feedstock fuel $e \in \mathbb{E}^{buy}_{a} $ bought from area $a$ in period $t$ for energy hub activation of Up/Down regulation service }

\nomenclature[D]{\(q^{{sell}^{+}}_{a,e,t}/q^{{sell}^{-}}_{a,e,t}\)}{Quantity of end product fuel $e \in \mathbb{E}^{sell}_{a}$ sold to area $a$ in period $t$ for energy hub activation of Up/Down regulation service}

\nomenclature[D]{\(f^{{flow}^{+}}_{a,a',e,t}/f^{{flow}^{-}}_{a,a',e,t}\)}{Flow from area $a$ to $a' $ of fuel $e$ in period $t$ for Up/Down regulation activity }

%



%
%

\nomenclature[D]{\(\lambda_{g,t}\)}{Binary indicating activity for providing Up or Down ancillary services of technology $g $ in period $t$}

\printnomenclature

\begin{equation}
\begin{aligned}
& \underset{\Xi
}{\operatorname{Maximize}}  
 \sum_{(a,e) \in \mathbb{E}^{buy}_{a}} \sum_{t} \pi^{sale}_{a,e,t} f^{sale}_{a,e,t} - \sum_{(a,e) \in \mathbb{E}^{sale}_{a}} \sum_{t}  \pi^{buy}_{a,e,t} f^{buy}_{a,e,t}   
\\&  
- \sum_{(g,e) \in \mathbb{E}^{def}_{g}} \sum_{t} x^{out}_{g,e,t} C^{var}_{g}  - \sum_{g \in \mathbb{G}^{UC}} \sum_{t}  c^{start}_{g,t}
\\& 
+ \sum_{g \in\mathbb{G}^{AS^{+}}}  \sum_{t} x^{cap^{+}}_{g,t} \pi^{+}_{t} + \sum_{g \in\mathbb{G}^{AS^{-}}}  \sum_{t} x^{cap^{-}}_{g,t} \pi^{-}_{t}
\\&
+\sum_{(a,e) \in \mathbb{E}^{buy}_{a}} \sum_{t} \pi^{sale}_{a,e,t} f^{sale^{+}}_{a,e,t} - \sum_{(a,e) \in \mathbb{E}^{sale}_{a}} \sum_{t}  \pi^{buy}_{a,e,t} f^{buy^{+}}_{a,e,t}   
\\&
-\sum_{(a,e) \in \mathbb{E}^{buy}_{a}} \sum_{t} \pi^{sale}_{a,e,t} f^{sale^{-}}_{a,e,t} + \sum_{(a,e) \in \mathbb{E}^{sale}_{a}} \sum_{t}  \pi^{buy}_{a,e,t} f^{buy^{-}}_{a,e,t} 
\\&
- \sum_{(g,e) \in (\mathbb{E}^{def}_{g} \cap \mathbb{G}^{AS^{+}})} \sum_{t} x^{out^{+}}_{g,e,t} C^{var}_{g}
+ \sum_{(g,e) \in (\mathbb{E}^{def}_{g} \cap \mathbb{G}^{AS^{-}})} \sum_{t} x^{out^{-}}_{g,e,t} C^{var}_{g}
\label{eq: AObjectivefunctionderPro}
\end{aligned}
\end{equation}


\begin{equation}
    \sigma^{in}_{g,e}  = \frac{F^{in}_{g,e}}{ \sum\limits_{e \in \mathbb{E}^{in}_{g}}F^{in}_{g,e}} {E}^{in}_{g} \quad \forall g \in \mathbb{G} ,e \in \mathbb{E}^{in}_{g}
    \label{eq:AInsharedefinied}
\end{equation}

\begin{equation}
    \sigma^{out}_{g,e}  = \frac{F^{out}_{g,e}}{ \sum\limits_{e \in \mathbb{E}^{out}_{g}}F^{out}_{g,e}}  \quad \forall g \in \mathbb{G} ,e \in \mathbb{E}^{out}_{g}
    \label{eq:AOutsharedefinied}
\end{equation}

\begin{equation}
    \Theta_{g}  = \frac{\sum\limits_{e \in \mathbb{E}^{in}_{g}}F^{in}_{g,e}}{ \sum\limits_{e \in \mathbb{E}^{out}_{g}}F^{out}_{g,e}}  \quad \forall g \in \mathbb{G} 
    \label{eq:APercentageDifference}
\end{equation}
\begin{equation}
    x^{in}_{g,e,t}  =  \sigma^{in}_{g,e} x^{total}_{g,t}  \quad \forall g \in \mathbb{G} ,e \in \mathbb{E}, t \in \mathbb{T}
    \label{eq:AFuelUse}
\end{equation}

\begin{equation}
    x^{out}_{g,e,t} = \sigma^{out}_{g,e} \Theta_{g} x^{total}_{g,t}    \quad \forall g \in  \mathbb{G} \backslash  ( \mathbb{G}^{S}  \cup \mathbb{G}^{LD} ) ,e \in \mathbb{E},t \in \mathbb{T}
    \label{eq:AGeneration}
\end{equation}
\begin{equation}
     x^{total}_{g,t} =   (\sum_{k \in K}  w^{SOS2}_{g,t,k} U_{g,k}) P^{max}_{g} \quad \forall g \in \mathbb{G}^{LD} e \in \mathbb{E}, t \in \mathbb{T}
    \label{eq:AImportFuelCurve}
\end{equation}
\begin{equation}
     x^{out}_{g,e,t} =  \sigma^{out}_{g,e} (\sum_{k \in K} w^{SOS2}_{g,t,k} U_{g,k} \Theta_{g,k}) P^{max}_{g} \quad \forall g \in \mathbb{G}^{LD}  ,e \in \mathbb{E},t \in \mathbb{T}
    \label{eq:AProductionFuelCurve}
\end{equation}
\begin{equation}
     \sum_{k \in K}  w^{SOS2}_{g,t,k} =  o_{g,t} \quad \forall g \in (\mathbb{G}^{LD} \cap  \mathbb{G}^{UC}) ,t \in \mathbb{T}
    \label{eq:AWeights}
\end{equation}
\begin{equation}
       x^{total}_{g,t} \leq P^{max}_{g} P^{profile}_{g,t}  \quad \forall g,t \in \mathbb{T} 
\label{eq:AlimitFuelusetot}
\end{equation}
\begin{equation}
       \sum_{e \in E} \left (x^{out}_{g,e,t} - x^{out}_{g,e,t-1} \right) \leq R^{up}_{g}  \quad \forall g \in  \mathbb{G}^{R} ,t \in \mathbb{T}
\label{eq:ARamp up no UC}
\end{equation}

\begin{equation}
       \sum_{e \in E} \left (x^{out}_{g,e,t-1} - x^{out}_{g,e,t} \right) \leq R^{down}_{g}  \quad \forall g \in  \mathbb{G}^{R} ,t \in T
\label{eq:ARamp Down no UC}
\end{equation}
\begin{equation}
      P^{min}_{g}o_{g,t}  \leq  x^{total}_{g,t}      \leq  P^{max}_{g}o^{}_{g,t}  \quad \forall g \in \mathbb{G}^{UC}, t \in \mathbb{T}
\label{eq:Amaxminproduction}
\end{equation}
\begin{equation}
        o_{g,t} - o_{g,t-1} \leq u_{g,t} \quad \forall g \in \mathbb{G}^{UC}, t \in \mathbb{T}
\label{eq:AStartupcostCondition}
\end{equation}

\begin{equation}
       c^{start}_{g,t} = C^{start}_{g}u_{g,t}    \quad \forall g \in \mathbb{G}^{UC}, t \in \mathbb{T}
\label{eq:AStartupcost}
\end{equation}
\begin{equation}
\begin{aligned}
       \sum_{e \in E} \left (x^{out}_{g,e,t} - x^{out}_{g,e,t-1} \right) \leq R^{up}_{g} o_{g,t-1} + P^{min}_{g} (1-o_{g,t-1})  \quad \forall g \in (\mathbb{G}^{UC} \cap \mathbb{G}^{R}),t \in \mathbb{T}
\label{eq:ARamp Up UC}
\end{aligned}
\end{equation}

\begin{equation}
\begin{aligned}
       \sum_{e \in E} \left (x^{out}_{g,e,t-1} - x^{out}_{g,e,t} \right) \leq R^{down}_{g} o_{g,t} + P^{min}_{g} (1-o_{g,t})  \quad \forall g \in (\mathbb{G}^{UC} \cap \mathbb{G}^{R}),t \in \mathbb{T}
\label{eq:ARamp Down UC}
\end{aligned}
\end{equation}
\begin{equation}
    soc_{g,t} = soc_{g,t-1} + \Theta_g x^{total}_{g,t} - \sum_{e \in E^{out}_{g}}  \sigma^{out}_{g,e} x^{out}_{g,e,t}   \quad \forall g \in \mathbb{G}^{s} ,e \in \mathbb{E}, t \in \mathbb{T}
    \label{eq:AProductionStorage}
\end{equation}

\begin{equation}
    soc_{g,t} \leq SOC^{max}_{g}    \quad \forall g \in \mathbb{G}^{s}, t \in \mathbb{T}
    \label{eq:Amax volume}
\end{equation}

\begin{equation}
    soc_{g,t=0} = soc_{g,t=T} = SOC^{init}_{g}   \quad \forall g \in \mathbb{G}^{s} 
    \label{eq:AvolinitLast}
\end{equation}
\begin{equation}
  P^{min}_{g} \nu_{g,t}  \leq x^{total}_{g,t} \leq P^{max}_{g} \nu_{g,t}    \quad \forall g \in \mathbb{G}^{s}, t \in \mathbb{T}
    \label{eq:Amax charging}
\end{equation}
\begin{equation}
 P^{min}_{g} (1-\nu_{g,t})   \leq \sum_{e \in E^{out}_{g}}  \sigma^{out}_{g,e} x^{out}_{g,e,t}  \leq P^{max}_{g} (1-\nu_{g,t})    \quad \forall g \in \mathbb{G}^{s}, t \in \mathbb{T}
    \label{eq:Amax discharging}
\end{equation}

\begin{equation}
\begin{aligned}
    & q^{buy}_{a,e,t} + \sum_{a' \in \mathbb{F}_{(a\prime,a)}} f^{flow}_{a\prime,a,e,t} + \sum_{g \in (\mathbb{G}_{a} \cap \mathbb{E}^{out}_{g})} x^{out}_{g,e,t} =     
   \\
   & q^{sale}_{a,e,t} + \sum_{a \in \mathbb{F}_{(a,a\prime)}} f^{flow}_{a,a\prime,e,t} + \sum_{g \in (\mathbb{G}_{a} \cap \mathbb{E}^{in}_{g})} x^{in}_{g,e,t}            \quad \forall  (a,a') \in \mathbb{A}, e \in \mathbb{E},t \in \mathbb{T}
    \label{eq:AEnergyBalance}
\end{aligned}
\end{equation}


\begin{equation}
  \sum_{a \in \mathbb{A}} \sum_{t} q^{buy}_{a,e,t} - \sum_{g \in (\mathbb{G}_{a} \cap \mathbb{E}^{out}_{g})} x^{out}_{g,e,t}   \leq \gamma  \sum_{g \in \mathbb{G}_{a}} \sum_{t}   x^{in}_{g,e,t} \quad \forall e \in \mathbb{E}=\{electricity \}
\label{eq:ALimitElectricityUse}
\end{equation}

\begin{equation}
    x^{in^{+}}_{g,e,t}  =   \sigma^{in}_{g,e} x^{total^{+}}_{g,t} \quad \forall g \in \mathbb{G}^{AS^{+}} ,e \in \mathbb{E}, t \in \mathbb{T}
    \label{eq:AFuelUse^{+}}
\end{equation}

\begin{equation}
    x^{out^{+}}_{g,e,t} =  \sigma^{out}_{g,e} \Theta_{g} x^{total^{+}}_{g,t}  \quad \forall g \in  \mathbb{G}^{AS^{+}} \backslash  ( \mathbb{G}^{S}  \cup \mathbb{G}^{LD} )   ,e \in \mathbb{E},t \in \mathbb{T}
    \label{eq:AGeneration^{-}}
\end{equation}

\begin{equation}
    x^{in^{-}}_{g,e,t}  =   \sigma^{in}_{g,e} x^{total^{-}}_{g,t} \quad \forall g \in \mathbb{G}^{AS^{-}} ,e \in \mathbb{E}, t \in \mathbb{T}
    \label{eq:AFuelUse^{-}}
\end{equation}

\begin{equation}
    x^{out^{-}}_{g,e,t} =  \sigma^{out}_{g,e} \Theta_{g} x^{total^{-}}_{g,t}  \quad \forall  g \in  \mathbb{G}^{AS^{-}} \backslash  ( \mathbb{G}^{S}  \cup \mathbb{G}^{LD} ) ,e \in \mathbb{E},t \in \mathbb{T}
    \label{eq:AGeneration^{-}}
\end{equation}

\begin{equation}
    x^{total^{+}}_{g,t} = \pi^{\omega}_{t} x^{cap^{+}}_{g,t}  \quad \forall g \in \mathbb{G}^{AS^{+}} ,e \in \mathbb{E}, t \in \mathbb{T}
    \label{eq:AFuelUsetotal^{+}}
\end{equation}

\begin{equation}
    x^{total^{-}}_{g,t} = \pi^{\omega}_{t} x^{cap^{-}}_{g,t}  \quad \forall g \in \mathbb{G}^{AS^{-}} ,e \in \mathbb{E}, t \in \mathbb{T}
    \label{eq:AFuelUsetotal^{-}}
\end{equation}

\begin{equation}
    x^{cap^{+}}_{g,t} \leq R^{+}_{g}P^{profile}_{g,t}\lambda_{g,t}  \quad \forall g \in \mathbb{G}^{AS^{+}} ,e \in \mathbb{E}, t \in \mathbb{T}
    \label{eq:AReservedCap^{+}}
\end{equation}

\begin{equation}
    x^{cap^{-}}_{g,t} \leq R^{-}_{g}P^{profile}_{g,t}(1-\lambda_{g,t})  \quad \forall g \in \mathbb{G}^{AS^{-}} ,e \in \mathbb{E}, t \in \mathbb{T}
    \label{eq:AReservedCap^{-}}
\end{equation}

\begin{equation}
    x^{total}_{g,t} \leq P^{max}_{g}P^{profile}_{g,t} - x^{cap^{+}}_{g,t}  \quad \forall g \in \mathbb{G}^{AS^{+}} ,e \in \mathbb{E}, t \in \mathbb{T}
    \label{eq:ACoupledReservedCap^{+}}
\end{equation}

\begin{equation}
    x^{total}_{g,t} \geq  x^{cap^{-}}_{g,t}  \quad \forall g \in \mathbb{G}^{AS^{-}} ,e \in \mathbb{E}, t \in \mathbb{T}
    \label{eq:ACoupledReservedCap^{-}}
\end{equation}

\begin{equation}
\begin{aligned}
       \sum_{e \in E} \left (x^{out}_{g,e,t} - x^{out}_{g,e,t-1} + x^{out^{+}}_{g,e,t} \right) \leq R^{up}_{g} o_{g,t-1} + P^{min}_{g} (1-o_{g,t-1}) 
       \\ 
       \quad \forall g \in (\mathbb{G^{UC}} \cap \mathbb{G}^{R}\cap \mathbb{G}^{AS^{+}}),t \in \mathbb{T}
\label{eq:ARamp Up UC_couplemarkets}
\end{aligned}
\end{equation}

\begin{equation}
\begin{aligned}
       \sum_{e \in E} \left (x^{out}_{g,e,t-1} - x^{out}_{g,e,t}
       - x^{out^{-}}_{g,e,t}\right) \leq R^{down}_{g} o^{online}_{g,t} + P^{min}_{g} (1-o_{g,t})  
       \\
       \quad \forall g \in (\mathbb{G}^{UC} \cap \mathbb{G}^{R}\cap \mathbb{G}^{AS^{-}}),t \in \mathbb{T}
\label{eq:ARamp Down UC_couplemarkets}
\end{aligned}
\end{equation}

\begin{equation}
       x^{total}_{g,t} + x^{cap^{+}}_{g,t}     \leq  P^{max}_{g}o^{}_{g,t}  \quad \forall g \in (\mathbb{G}^{UC} \cap \mathbb{G}^{AS^{+}}), t \in \mathbb{T}
\label{eq:Amax production_couplemarkets}
\end{equation}

\begin{equation}
       x^{total}_{g,t}  + x^{cap^{+}}_{g,t}    \geq  P^{min}_{g}o_{g,t}  \quad \forall g \in (\mathbb{G}^{UC} \cap \mathbb{G}^{AS^{+}}), t \in \mathbb{T}
\label{eq:Aminproduction_couplemarkets}
\end{equation}

\begin{equation}
\begin{aligned}
    q^{buy^{+}}_{a,e,t} + \sum_{a' \in \mathbb{F}_{(a,a)}} f^{flow^{+}}_{a',a,e,t} + \sum_{g \in (\mathbb{G}_{a} \cap \mathbb{G}^{AS^{+}}\cap \mathbb{E}^{in}_{g})} x^{out^{+}}_{g,e,t}  =     q^{sale^{+}}_{a,e,t} + \sum_{a \in \mathbb{F}_{(a,a)}} f^{flow^{+}}_{a,a',e,t}
   \\
   + \sum_{g \in (\mathbb{G}_{a} \cap \mathbb{G}^{AS^{+}} \cap \mathbb{E}^{out}_{g})} x^{in^{+}}_{g,e,t}          \quad \forall  (a,a') \in \mathbb{A}, e \in \mathbb{E},t \in \mathbb{T}
    \label{eq:AEnergyBalanceUP}
\end{aligned}
\end{equation}

\begin{equation}
\begin{aligned}
    q^{buy^{-}}_{a,e,t} + \sum_{a' \in \mathbb{F}_{(a,a)}} f^{flow^{-}}_{a',a,e,t} + \sum_{g \in (\mathbb{G}_{a} \cap \mathbb{G}^{AS^{-}}\cap \mathbb{E}^{in}_{g})} x^{out^{+}}_{g,e,t}  =     q^{sale^{-}}_{a,e,t} + \sum_{a \in \mathbb{F}_{(a,a)}} f^{flow^{-}}_{a,a',e,t}
   \\
   + \sum_{g \in (\mathbb{G}_{a}\cap \mathbb{G}^{AS^{-}} \cap \mathbb{E}^{out}_{g})} x^{in^{-}}_{g,e,t}          \quad \forall  (a,a') \in \mathbb{A}, e \in \mathbb{E},t \in \mathbb{T}
    \label{eq:AEnergyBalanceDown}
\end{aligned}
\end{equation}

\begin{equation}
q^{sale}_{a,e,t} + q^{sale^{+}}_{a,e,t} - q^{sale^{-}}_{a,e,t} = D_{a,e,t} \quad \forall a \in \mathbb{A}, e \in \mathbb{E}^{D},  t \in \mathbb{T}
\label{eq:ADemand Balancefor all markets}
\end{equation}

\begin{equation}
       f^{flow}_{a,a\prime,e,t} + f^{flow^{+}}_{a,a\prime,e,t}\leq X^{cap}_{a,a\prime,e,t} \quad \forall (a,a\prime) \in \mathbb{A} , e \in \mathbb{E} , t \in \mathbb{T}
\label{eq:AMaxinterconectionflowsallmarkets}
\end{equation}

\begin{equation}
q^{buy^{+}}_{a,e,t} \leq \sum_{g \in (\mathbb{E}^{In}_{g} \cap \mathbb{G}^{AS^{+}})} x^{in^{+}}_{g,e,t}  \quad \forall a \in \mathbb{A}, e \in \mathbb{E}, t \in \mathbb{T}
\label{eq:AMaxBuyUp}
\end{equation}

\begin{equation}
q^{buy^{-}}_{a,e,t} \leq q^{buy}_{a,e,t}   \quad \forall a \in \mathbb{A}, e \in \mathbb{E}, t \in \mathbb{T}
\label{eq:AMaxBuydown}
\end{equation}

\begin{equation}
q^{sale^{+}}_{a,e,t} \leq \sum_{g \in (\mathbb{E}^{In}_{g} \cap \mathbb{G}^{AS^{+}})} x^{out^{+}}_{g,e,t}  \quad \forall a \in \mathbb{A}, e \in \mathbb{E}, t \in \mathbb{T}
\label{eq:AMaxSaleUp}
\end{equation}

\begin{equation}
q^{sale^{-}}_{a,e,t} \leq q^{sale}_{a,e,t}   \quad \forall a \in \mathbb{A}, e \in \mathbb{E}, t \in \mathbb{T}
\label{eq:AMaxSaledown}
\end{equation}

\begin{equation}
\begin{aligned}
    soc_{g,t} = soc_{g,t-1} + x^{total}_{g,t}\Theta_g - \sum_{e \in E^{out}_{g}}  x^{out}_{g,e,t} \sigma^{out}_{g,e}   
    + x^{total^{+}}_{g,t}\Theta_g 
    \\
     - \sum_{e \in E^{out}_{g}}  x^{out^{+}}_{g,e,t} \sigma^{out}_{g,e} - x^{total^{-}}_{g,t}\Theta_g + \sum_{e \in E^{out}_{g}}  x^{out^{-}}_{g,e,t} \sigma^{out}_{g,e}
    \\
    \quad \forall g \in (\mathbb{G}^{s}  \cap \mathbb{G}^{AS}) ,e \in \mathbb{E}, t \in \mathbb{T} /t=1
    \label{eq:AProductionStorage_coupledmarkets}
\end{aligned}
\end{equation}

\begin{equation}
    x^{total}_{g,t} + x^{cap^{+}}_{g,t} - x^{cap^{-}}_{g,t} \leq P^{max}_{g} \nu_{g,t}    \quad \forall  g \in (\mathbb{G}^{s}  \cap \mathbb{G}^{AS}), t \in \mathbb{T}
    \label{eq:Amax charging_couplemarkets}
\end{equation}

\begin{equation}
    x^{total}_{g,t} + x^{cap^{+}}_{g,t} - x^{cap^{-}}_{g,t} \geq P^{min}_{g} \nu_{g,t}    \quad \forall  g \in (G^{s}  \cap \mathbb{G}^{AS}), t \in \mathbb{T}
    \label{eq:Amin charging_couplemarkets_couplemarkets}
\end{equation}

\begin{equation}
\begin{aligned}
    \sum_{e \in E^{out}_{g}} ( x^{out}_{g,e,t} + x^{out^{+}}_{g,e,t} - x^{out^{+}}_{g,e,t} ) \sigma^{out}_{g,e}  \leq P^{max}_{g} (1-\nu_{g,t})   
    \\
    \quad \forall  g \in (G^{s}  \cap \mathbb{G}^{AS}), t \in \mathbb{T}
    \label{eq:Amax discharging_couplemarkets}
\end{aligned}
\end{equation}

\begin{equation}
\begin{aligned}
    \sum_{e \in E^{out}_{g}}  ( x^{out}_{g,e,t} + x^{out^{+}}_{g,e,t} - x^{out^{+}}_{g,e,t} ) \sigma^{out}_{g,e}  \geq P^{min}_{g} (1-\nu_{g,t})  
    \\
    \quad \forall  g \in (G^{s}  \cap \mathbb{G}^{AS}), t \in \mathbb{T}
    \label{eq:Amax discharging_couplemarkets}
\end{aligned}
\end{equation}





\end{document}